\documentclass{article}
\usepackage{amsmath,epsfig}
\usepackage{bm}
\usepackage{algorithm}
\usepackage{algorithmic}
\usepackage{subfigure}
\usepackage{color}
\usepackage{comment}
\usepackage[english]{babel}

\newcommand{\MP}{ \mbox{Pr} }
\newcommand{\MV}{ \mbox{Var} }
\newcommand{\SD}{ {\cal D} }

\newcommand{\Sr}{ {\mathbf r} }
\newcommand{\SA}{ {\cal A}}
\newcommand{\Sa}{ {\widetilde \alpha} }

\newtheorem{lemma}{{\bf Lemma}}
\newtheorem{theorem}{{\bf Theorem}}
\newtheorem{proposition}{{\bf Proposition}}

\def\done{\hspace*{\fill} \rule{1.8mm}{2.5mm} \\}

\title{Mathematical Modeling of Product Rating: Sufficiency, Misbehavior and Aggregation Rules}

\author{
Hong Xie \hspace{0.1in} John C.S. Lui\\
Computer Science \& Engineering Department\\
The Chinese University of Hong Kong\\
Email: \{hxie,cslui\}@cse.cuhk.edu.hk
}

\begin{document}
\maketitle

\begin{abstract}
Many web services like eBay, Tripadvisor, Epinions, etc, provide historical product
ratings so that users can evaluate the quality of products. Product ratings are
important since they affect how well a product will be adopted by the market.
The challenge is that we only have
{\em "partial information"} on these ratings:
Each user provides ratings to only a
"{\em small subset of products}".  Under this partial information setting,
we explore a number of fundamental questions: What is the "{\em
minimum number of ratings}" a product needs so one can make
a reliable evaluation of its quality? How users' {\em misbehavior}
(such as {\em cheating}) in product rating may affect the
evaluation result? To answer these questions, we present a formal
mathematical model of product evaluation based on partial
information.  We derive theoretical bounds on the
minimum number of ratings needed to produce a reliable indicator
of a product's quality.  We also extend our model to accommodate users' misbehavior
in product rating.
We carry out experiments using both synthetic and real-world data (from TripAdvisor,
Amazon and eBay) to validate our model, and also show that using the
"majority rating rule" to aggregate product ratings, it produces more reliable
and robust product evaluation results than the "average rating rule".
\end{abstract}


\section{ Introduction}
\label{section:introduction}

Nowadays, it is common in many web services that users can contribute
their opinions in the form of ratings or reviews.  For example, we see
ratings or reviews in content sharing sites (i.e., Flickr, YouTube, etc),
online recommendation systems (i.e., Amazon, MovieLens, etc),
product review sites (i.e., Epinions, TripAdvisor, etc),
and online e-commerce systems (i.e., eBay, etc), etc.
With  these ratings and product reviews, one can perform information search or
make purchase decisions by taking advantage of the opinions of other users
(aka "wisdom of the crowd").
Web-based online rating systems can be briefly described as follows:
There are a number of items (either products or services), each user provides
ratings or reviews to a {\em subset of items}, and ratings of
an item will be available to the public.

Based on how these ratings are being used, online rating systems can be classified into
two categories.  The first category interprets ratings as public
assessment on products\cite{ref:rating_agg_ors, ref:factor_based_rating_agg_ors}, where
the main interest is in assisting users to evaluate products' quality.
For example, in eBay, ratings are used to reflect the quality a product or the
reputation of a seller. Other examples of this type of web services include Tripadvisor,
Epinions, Wikipedia, etc.  For these web services, users want to know whether the quality
of a product is as good as its declared historical ratings, or whether the reputation
of a seller is fully reflected by its historical ratings.
The second category interprets ratings as users' preference information and they are
used in users' preference modeling.  This is the view adopted by researchers in
online recommendation systems\cite{ref:recsys_survey,rec_sys}, i.e.,
in Amazon and MovieLens,  users' ratings are used to infer users'
preferences\cite{rec_sys_eval_coll_fil_algo,grouplens_1994,rec_sys}
as to make personalized recommendations.

In this paper, we focus on the first category of online rating systems,
in which ratings are interpreted as product quality assessment.
Although there are wide deployment of product ratings in web services,
people only have {\em partial information} on these ratings: each user only expresses
ratings to a {\em small subset} of products.  Hence, it is important
to understand the "{\em accuracy}" and "{\em effectiveness}" of online rating systems.
Little attention has been paid to this fundamental question.
The goal of this paper is to fill this void. We first provide a
mathematical model for a general online rating system, then we
explore a number of fundamental questions,
i.e., (a) {\em What is the minimum number of ratings a product needs so as
to have a reliable reflection on its quality}?
(b) {\em How users' misbehavior may affect the the accuracy}?  To
the best of our knowledge, this is the first paper that provides a formal
mathematical model and analysis on such problem.  Our contributions are:
\begin{itemize}
\item We propose a mathematical model to specify the
       rating behavior of users to a set of products.  Our model
       is general enough to represent both honest and misbehaving
       users in product rating.  We analyze the model and derive
       the minimum number of ratings we need to produce a reliable
       reflection on a product's quality. We also derive the
       minimum fraction of misbehaving users such that they can
       manipulate an online rating system {\em (please refer to
       Section \ref{section:model} and
       \ref{section: analysis})}.
\item We propose an inference algorithm to infer parameters of
       our model from {\em partial information}, say available
       products' ratings {\em (please refer to Section \ref{section:infer_algo})}.
\item We show that the {\em majority rule} is more robust and
       insensitive to users' misbehavior as compared with the
       {\em average scoring rule} {\em (please refer to
       Section \ref{section: analysis} and Section \ref{section:exp_syn_data})}.
\item We perform experiments using both synthetic data and
       real-world data (data set from TripAdvisor, Amazon and eBay)
       to validate our framework and to
       examine various factors that may affect product quality evaluation.
       We find a number of interesting observations,
       i.e., the system can resist a small fraction of misbehaving
       users. For TtripAdvisor, Amazon, and Ebay, we also show why the {\em majority rule} is more
       robust and reliable than the {\em average scoring rule} in
       evaluating products' quality.  {\em (please refer to
       Section \ref{section:exp_syn_data} and Section \ref{section:exp_real_data})}.
\end{itemize}

This is the outline of the paper. In \S\ref{section:model}, we present the mathematical
model of product rating systems.  In \S\ref{section: analysis}, we show the analysis
and derive theoretical results for two rating aggregation rules under normal or
misbehaving setting.  In \S \ref{section:infer_algo}, we present an inference
algorithm to infer the parameters of our model.
In \S \ref{section:exp_syn_data}, we present our experimental results using
synthetic data. In \S \ref{section:exp_real_data}, we present our experimental
results using real-world data. Related work is given
in \S \ref{section:related_work} and \S \ref{section:conclusion} concludes.

\section{ Mathematical Model}
\label{section:model}

Let there be a finite set of $N$ products denoted by $P_1, \ldots, P_N$.
We have $M$ users denote by $U_1, \ldots, U_M$.
Each user only expresses ratings to a {\em subset} of products on an $m$-level
cardinal metric where ratings are drawn from $\{1, \ldots, m\}$, i.e.,
a 3-level cardinal metric could be: $\{1 \!=\! \mbox{"poor"},
2 \!=\! \mbox{"good"}, 3 \!=\! \mbox{"excellent"}\}$.  Ratings from users are
independent. Product $P_i$ is rated by $n_i \! \leq\! M$ users.
Let $\Sr_i \!=\! \{r_{i,1}, \ldots, r_{i,M}\}$ denote a set of $M$ ratings of
product $P_i$, where $r_{i,j} \!\in\! \{1, \ldots, m\}$ if user $U_j$ rates
product $P_i$, otherwise $r_{i,j} \!=\! 0$. Higher rating implies higher quality.
We treat the available ratings as {\em partial information} because some
$r_{i,k}\!=\!0$.  To make a purchase decision, a user evaluates the quality of
products relying on historical product ratings.
The objective of this work is to answer how various factors can
influence the evaluation accuracy.

Let $\SA$ denote a rating aggregation rule which is used to summarize all
ratings of a product. Two commonly used rating aggregating rules are the
{\em majority rule} and the {\em average scoring rule}.

\noindent {\bf $\bullet$ Majority rule.}
It evaluates the quality of a product via the {\em majority} ratings.
Let $\ell_i \!\in\! \{1, \ldots, m\}$ represents the label which reflects the
{\em true} quality of product $P_i$ under this policy.
Let $\widehat{\ell}_i$ denote the {\em evaluated label}
of product $P_i$ produced by applying the {\em majority rule}.
The label $\widehat{\ell}_i$ can be computed as follows. Let $n_{i,k}$ be the
{\em number} of ratings of $P_i$ that is of level $k$, where $k \!\in\! \{1,...,m\}$.
Then $\widehat{\ell}_i \!=\! \arg\max_{k} \{n_{i,k}\}$.
There are a number of interesting questions to explore, i.e.,
How many ratings,
or $n_i$, do we need so as to have a strong guarantee in revealing the true label,
or $\MP \left[ \widehat{\ell}_i \!=\! \ell_i\right]$ with high probability?

\noindent {\bf $\bullet$ Average scoring rule.}  It evaluates the
quality of a product via averaging all ratings of that product.
Let $\gamma_i \!\in\! [1, m]$ denote the {\em true} quality of $P_i$
under this policy.
Let $\widehat{r}_i \!=\! \sum_j r_{i,j} / n_i$ denote the {\em evaluated label}
of $P_i$ using the average scoring rule.
Again, there are number of
interesting questions to explore, i.e., how many ratings, or $n_i$ do we need,
such that $\widehat{r}_i$ accurately reflects $\gamma_i$?

\noindent {\bf Remark:} There should be no confusion between
$\ell_i$ and $\gamma_i$. They show two different ways in
evaluating the quality of products either by the ratings of the
majority population, or by averaging the ratings on the whole population respectively.

Because each user may rate a product differently.
To describe a user's rating behavior, let us
first present the probabilistic model in product rating.

\subsection{ Model for rating behavior }

When rating a product, a user needs to evaluate its quality.  We
consider two most important factors that may affect the evaluation:
(a) intrinsic quality of products, and (b) preference of users on a product.
The rating behavior of different users can be modeled by a random variable
in which one can vary its mean or variance to reflect the above factors.
Specifically, a higher mean implies that the product has a higher intrinsic quality,
while a smaller variance implies that a user prefers a particular branding.

To illustrate, consider the user $U_j$ who rates product $P_i$, and
the rating is $r_{i,j}$. The rating $r_{i,j}$ is a random variable,
and its probability mass function (pmf) is:
\begin{equation}
    \MP [r_{i,j} = k ] = \rho_k,
    \hspace{0.1 in} k = 1, \ldots, m,
\end{equation}
where $\rho_k \geq 0$ and $\sum_{k=1}^m \rho_k \!=\! 1$. One can
set the mean or variance by varying the probability mass distribution $\{
\rho_1, \ldots, \rho_m \}$ to reflect different intrinsic quality
or users preference on product $P_i$.
To model the collective rating behavior of the whole user population over
product $P_i$, let
\[
    \mathcal{S} = \left\{(\theta_1, \ldots, \theta_m) |
          \sum\nolimits_{i=1}^m \theta_i = 1, \theta_i \geq 0, \forall i \right\}
\]
denote the space over all the possible probability distributions of
a rating. We assume that there is an underlying distribution, say
$\SD (P_i)$, over the space $\mathcal{S}$, that defines the
collective rating behavior of the whole user population over product
$P_i$. In this study, $\SD(P_i)$ is a Dirichlet distribution
$Dirichlet( {\bm \alpha}_i)$ with density function:
\begin{equation}
    p ( \theta_1, \ldots, \theta_m) = \frac{\Gamma (\sum_{k=1}^m \alpha_{i,k})}{\prod_{k=1}^m \Gamma (\alpha_{i,k})}
    \prod_{k=1}^m \theta_{k}^{\alpha_{i,k} - 1}.
\end{equation}
where ${\bm \alpha}_i \!=\! (\alpha_{i,1}, \ldots, \alpha_{i, m})$,
$\sum_{k=1}^{m} \alpha_{i,k} \!=\! 1$ and $\alpha_{i,k} > 0, \forall
k$.

\noindent
{\bf Remark:} The choice of Dirichlet distribution is validated in \cite{ref:prml}.

\section{Theoretical Analysis}  \label{section: analysis}

In this section, we analyze two most representative rating aggregation rules:
{\em majority rule} and {\em average scoring rule}, that are widely used in
online rating systems. The central question we seek to answer is how many ratings
of a given product we need so that the aggregate rating is statistically accurate
to reflect the ground truth of that product's quality? Furthermore, we also
analyze the impact of misbehavior in product rating, e.g., when users intentionally
rate a product beyond its ground truth level. In this section, we
assume that the model parameters $\bm \alpha_i, \forall i$, are given.
In Section \ref{section:infer_algo}, we will show how to infer
these parameters from historical ratings.

\subsection{ Majority rule}

{\em Majority rule} is suggested as a good rule to aggregate
ratings\cite{web:trip_advisor}.  This rule helps a user to understand how receptive
the majority of the masses are toward a given product.  Note that we only have
{\em partial information}, and this makes it challenging to extract the true label of
products.  There are a number of interesting questions to explore, i.e., how many
ratings do we need so that we can extract the true label with high probability?
What is the effect of misbehavior in ratings? Let us begin our exploration by a
simple case, say all users rate products honestly.

\noindent {\bf Analysis for Honest Rating:}  In this case, we assume that with a
large enough number of ratings, one can successfully extract the true label of products.
Recall that $\widehat{\ell}_i$ is the label of product $P_i$ produced by using the
{\em majority rule} on its ratings, and $\ell_i$ is the true label of product $P_i$.
Intuitively, the true label of product $P_i$ occurs when
$\ell_i \!=\! \lim_{n_i \rightarrow \infty} \widehat{\ell}_i$. Recall that $P_i$ has
$n_i$ ratings and its rating set is $\Sr_i$, where missing ratings are denoted by zeros.
For convenience, here we use $\Sr ^{+}_i \!=\! \{r^+_{i,1}, \ldots, r^+_{i, n_i}\}$
to denote a set of all positive (or observed) ratings of $\Sr_i$.  Let us state the
probability mass function (pmf) of $r^+_{i,j}$, $j \!=\! 1, \ldots, n_i$ in the
following lemma.

\begin{lemma}
{\em The pmf of the rating $r^+_{i,j}$ is:
\[
    \MP \left[ r^+_{i,j} = k \right] = \alpha_{i,k},
    \hspace{0.2 in}
    \mbox{for $k = 1, \ldots, m$},
\]
where $i \!=\! 1,\ldots, N $ and  $j \!=\! 1, \ldots, n_i$.
\label{lemma:rating_dis} }
\end{lemma}

\noindent {\bf Proof:}  Please refer to the appendix for derivation. \done

Lemma \ref{lemma:rating_dis} states that the probability of product $P_i$ receiving a
rating $k$ is $\alpha_{i,k}$. It follows that
$\ell_i \!=\! \lim_{n_i \rightarrow \infty} \widehat{\ell}_i \!=\! \arg\max_{k }
\{ \alpha_{i, k} \}$.  Let \mbox{$ \widetilde{\alpha}_i$} denote the second largest
value among $\alpha_{i,1}, \ldots, \alpha_{i, m}$.
Now we state the main theorem for the honest rating case.
\begin{theorem}
[Honest Rating Case] {\em Suppose all users rate honestly, and
product $P_i$ has at least
\begin{equation}
   n_i \geq n'_i = \frac{12 \alpha_{i,\ell_i} }{(\alpha_{i,\ell_i}
      - \widetilde{\alpha}_i)^2}\ln \frac{m}{\delta}
\label{ineq:mr_sincere:rating_bound}
\end{equation}
ratings, then one can claim with high probability that the true
label of product $P_i$ is $\widehat{\ell}_i$.  Mathematically, we
have:  $\MP \left[ \widehat{\ell}_i = \ell_i \right] \geq 1 -
\delta$. \label{theorem:mr_sincere} }
\end{theorem}

\noindent {\bf Proof:} Please refer to the appendix for derivation.
\done

Table \ref{table:mr_Sincere} shows some numerical results on the
minimum number of ratings we need when all users rate honestly.
It depicts the level of rating metric $m$,
the model parameter $\bm \alpha_i$, the success probability $1 \!-\!
\delta$ and the lower bound of the minimum number of ratings denoted
by $n'_i$ respectively.  From the table, we see that if we increase
the success probability $(1\!-\!\delta)$, we also need to increase $n'_i$.

{
\renewcommand{\arraystretch}{1.4}
\begin{table}[htb]
\centering {\renewcommand{\tabcolsep}{0.12cm}
{\small
\begin{tabular}{|c|c|c|c|c|}
\hline
$m$ &  ${\bm \alpha}_i=(\alpha_{i,1},\alpha_{i,2},\alpha_{i,3},\alpha_{i,4},
\alpha_{i,5}) $ & $ 1- \delta $ & $ n'_i $ \\ \hline \hline
5 & $(\frac{4}{35}, \frac{25}{35}, \frac{3}{35}, \frac{2}{35}, \frac{1}{35})$  & 0.7 & 67  \\ \hline
5 & $(\frac{4}{35}, \frac{25}{35}, \frac{3}{35}, \frac{2}{35}, \frac{1}{35})$  & 0.8 & 77  \\  \hline
5 & $(\frac{4}{35}, \frac{25}{35}, \frac{3}{35}, \frac{2}{35}, \frac{1}{35})$  & 0.9 & 93  \\  \hline
\end{tabular}
}
} \caption{Honest rating: minimum number of ratings.}
\label{table:mr_Sincere}
\end{table}
}

\noindent {\bf Remark:} Theorem \ref{theorem:mr_sincere} gives an lower bound on
the minimum number of ratings $n_i$. One may ask whether this bound is
sufficiently tight. The following theorem answers this question.

\begin{theorem}
[Tightness] {\em Suppose all users rate honestly
and $\widetilde{\alpha}_i \geq \frac{100}{101} \alpha_{i, \ell_i}$.
Then there exist a positive constant $\eta$, such that for any $\delta \leq \eta$,
if product $P_i$ has at most
\[
   n_i =  O \left(\frac{\alpha_{i, \ell_i}}{( \alpha_{i, \ell_i} -
   \widetilde{\alpha}_i )^2} \ln \frac{1}{\delta} \right)
\]
ratings, then we will fail to extract the true label with
probability at least $\Omega ( \delta )$.
\label{theorem:mr_sincere:tightness} }
\end{theorem}
{\bf Proof:} Please refer to the appendix for derivation. \done

\noindent {\bf Remark:} Theorem \ref{theorem:mr_sincere:tightness} states the
lower bound derived in Theorem \ref{theorem:mr_sincere}
is asymptotically tight in general.

\noindent {\bf Analysis of Rating under Misbehavior:}
Let us now explore the effect misbehavior in ratings. We consider the following typical
cases of misbehavior:
\begin{itemize}
   \item {\bf Random misbehavior.} A random misbehavior implies
       that a user assigns a random rating to a product.  This is
       one typical misbehavior, because sometimes a user may not
       want to spend time to evaluate the quality of a product so
       that user just generates random rating.
   \item {\bf Biased misbehavior:}  A biased misbehavior implies
       that a user is biased toward one particular rating, i.e., a
       user may be hired by a company to assign the lowest rating to a
       competitor's product, or
       assign the highest rating to his employer's product.  This type of
       misbehavior has been observed in many web services, i.e., reported in
       TripAdvisor\cite{web:trip_advisor}.
\end{itemize}

In the remaining of this sub-section, we {\em model} the ratings of
misbehaving users as noise.  We specify the collective rating
behavior of honest users to product $P_i$ using our model with
parameter $\bm \alpha_i$.  Again, $\ell_i \!=\!
\arg\max\{\alpha_{i,k}\}$ is the true label of product $P_i$ and
$\widetilde{\alpha}_i$ is the second largest value among
$\alpha_{i,1}, \ldots, \alpha_{i, m}$. The following theorem states
the effect of random misbehavior.

\begin{theorem}
[Resist Random Misbehavior] {\em Let $f_i < 1$ denote the fraction of
random misbehaving users who rate product $P_i$. If product $P_i$
has
\[
n_i \geq n'_i = \frac{ 12 (f_i / m + ( 1 - f_i ) \alpha_{i, \ell_i}) }
              { (1 - f_i)^2(\alpha_{i,\ell_i} - \widetilde{\alpha}_i )^2 }
          \ln \frac{m}{\delta}
\]
ratings, then one can claim with high probability that the true
label of product $P_i$ is $\widehat{\ell}_i$.  Mathematically, we
have: $ \MP\left[\widehat{\ell}_i = \ell_i \right] \geq 1 - \delta$.
} \label{theorem:mr_random_misb}
\end{theorem}

\noindent {\bf Proof:} Please refer to the appendix for derivation. \done

\noindent {\bf Remark:}
Theorem \ref{theorem:mr_random_misb} states that if the fraction of random misbehavior
is less than 1, then one can always extract the true label of a product  with a
large enough number of ratings.

Table \ref{table:examp_random_misb} shows some numerical results of the minimum number
of ratings when some ratings are assigned by random misbehaving users.
Table \ref{table:examp_random_misb} depicts $m$,
$\bm \alpha_i$, $1 \!-\! \delta$, $n'_i$, and the fraction of random misbehaving users
$f_i$ respectively. We can see that if we increase $f_i$, we also need to
increase the minimum number of ratings.

{
\renewcommand{\arraystretch}{1.4}
\begin{table}[htb]
\centering {\renewcommand{\tabcolsep}{0.12cm}
{\small
\begin{tabular}{|c|c|c|c|c|}
\hline
$m$ &  ${\bm \alpha}_i=(\alpha_{i,1},\alpha_{i,2},\alpha_{i,3},\alpha_{i,4},
\alpha_{i,5}) $ & $ 1- \delta $ & $f_i$ & $ n'_i $ \\ \hline \hline
5  & $(\frac{4}{35}, \frac{25}{35}, \frac{3}{35}, \frac{2}{35}, \frac{1}{35})$  & 0.8 & 0 & 77  \\  \hline
5  & $(\frac{4}{35}, \frac{25}{35}, \frac{3}{35}, \frac{2}{35}, \frac{1}{35})$  & 0.8 & 0.1 & 88  \\  \hline
5  & $(\frac{4}{35}, \frac{25}{35}, \frac{3}{35}, \frac{2}{35}, \frac{1}{35})$  & 0.8 & 0.2& 102  \\  \hline
\end{tabular}
}
} \caption{Random misbehavior: minimum number of ratings. }
\label{table:examp_random_misb}
\end{table}
}

We now consider the effect of biased misbehavior.  Let $\ell'_i \in
\{1, \ldots, m\}$ denote the rating that biased misbehaving users
bias toward. The goal of the biased users is to make the extracted
label of a product to be the label that they bias toward. Then, how
many biased users is needed such that they will achieve thier goal with high
probability? The following theorem answers this question.

\begin{theorem}
[Biased Misbehaving Users Win] {\em Let $f'_i $ denote the fraction
of biased misbehaving users who rate product $P_i$.  If the fraction
of biased misbehaving users is no less than $ f'_i >
(\alpha_{i,\ell_i} - \alpha_{i, \ell'_i} ) / ( 1 + \alpha_{i,\ell_i}
- \alpha_{i, \ell'_i}), $ then with at least
\[
n_i \geq \left\{
  \begin{aligned}
     & \frac{ 12 \left( f'_i + ( 1 - f'_i ) \alpha_{i, \ell'_i} \right) }
      { \left( f'_i + ( 1 - f'_i )( \alpha_{i, \ell'_i} - \alpha_{i, \ell_i} ) \right)^2 }
       \ln \frac{ m }{ \delta },  && \hspace{0.1 in} \ell'_i \neq \ell_i
       \\
     & \frac{ 12 \left( f'_i + ( 1 - f'_i ) \alpha_{i, \ell_i} \right) }
        { \left( f'_i+( 1 - f'_i )(\alpha_{i, \ell_i}-\widetilde{\alpha}_{i} ) \right)^2 }
       \ln \frac{ m }{ \delta }, && \hspace{0.1 in} \ell'_i = \ell_i
  \end{aligned}
    \right.
\]
ratings, one can claim with high probability that the biased
misbehaving users win.  Mathematically, we have: $ \MP \left[
\widehat{\ell}_i = \ell'_i \right] \geq 1 - \delta$.
\label{theorem:mr_bias:bias_win} }
\end{theorem}

\noindent {\bf Proof:} Please refer to the appendix for derivation. \done

\noindent{\bf Remark:} Theorem \ref{theorem:mr_bias:bias_win} states
two conditions for biased misbehaving users to win: a large enough
fraction of biased misbehaving users and a large enough number of
such ratings.

Table \ref{table:examp_frac_biasd_win} shows some numerical results of the
minimum fraction of biased misbehaving users so that they can distort the
evaluation process. Table \ref{table:examp_frac_biasd_win} depicts
$m$, $\bm \alpha_i$, $n'_i$, and the label that the biased misbehaving users bias
toward $\ell'_i$ respectively.  From Table \ref{table:examp_frac_biasd_win} we
could see that if biased misbehaving users biased toward 5 (or 1), and to ensure
they can distort, the minimum fraction of biased misbehaving users
we need is 0.407 (or 0.375).

{
\renewcommand{\arraystretch}{1.4}
\begin{table}[htb]
\centering {\renewcommand{\tabcolsep}{0.12cm}
{\small
\begin{tabular}{|c|c|c|c|}
\hline
$m$  & ${\bm \alpha}_i=(\alpha_{i,1},\alpha_{i,2},\alpha_{i,3},\alpha_{i,4},
\alpha_{i,5})$ & $ \ell'_i $ & $f'_i$
\\ \hline \hline
5  & $(\frac{4}{35}, \frac{25}{35}, \frac{3}{35}, \frac{2}{35}, \frac{1}{35})$  & 5 & 0.407   \\  \hline
5  & $(\frac{4}{35}, \frac{25}{35}, \frac{3}{35}, \frac{2}{35}, \frac{1}{35})$  & 1 & 0.375   \\  \hline
\end{tabular}
}
}
\caption{Biased misbehavior: minimum fraction of misbehaving users
to distort the evaluation.}
\label{table:examp_frac_biasd_win}
\end{table}
} \vspace{-0.1 in}

\begin{proposition}
{\em Suppose that the misbehaving users bias against the
ground truth label, or $\ell'_i \neq \ell_i$.  If the fraction of
biased misbehaving users who rate product $P_i$ is at least $ f'_i >
( \alpha_{i,\ell_i} - \alpha_{i, \ell'_i} ) / ( 1 +
\alpha_{i,\ell_i} - \alpha_{i, \ell'_i} ) $, then it is impossible
to extract the true label with high probability no matter how many
ratings we have. }
\end{proposition}

The following theorem states that when the fraction of biased
misbehaving users is small, we can resist this type of anomaly.

\begin{theorem}
[Resist Biased Misbehavior] {\em Suppose that the biased misbehaving users
bias against the ground truth label, say $\ell'_i \neq \ell_i$.  If
the fraction of biased misbehaving users who rate product $P_i$ is
at most $ f'_i < ( \alpha_{i,\ell_i} - \alpha_{i, \ell'_i} ) / ( 1 +
\alpha_{i,\ell_i} - \alpha_{i, \ell'_i} )$, then with at least
\[
n_i \geq n'_i = \frac{ 12 ( 1 - f'_i ) \alpha_{i, \ell_i}  \ln \frac{ m }{ \delta } }
  { \left( ( 1 \!-\! f'_i )\alpha_{i, \ell_i} \!-\! \mbox{max}\left\{f'_i \!+\!
        (1 \!-\! f'_i)\alpha_{i,\ell'_i}, (1 \!-\! f'_i)\widetilde{\alpha}_i \right\}
    \right)^2
  }
\]
ratings, then one can claim with high probability that the true
label of product $P_i$ is $\widehat{\ell}_i$.  Mathematically, we
have: $ \MP \left[ \widehat{\ell}_i \neq \ell_i \right] \geq 1 -
\delta $. \label{theorem:mr_bias:extract_true_label} }
\end{theorem}
{\bf Proof:} The derivation is similar to Theorem \ref{theorem:mr_bias:bias_win}. \done

Table \ref{table:examp_bias_misb} shows some numerical examples of
the minimum number of ratings we need to resist biased misbehavior,
when the fraction of biased misbehaving users
is small.  From Table \ref{table:examp_bias_misb}, we could
see that when we increase the fraction of biased
misbehaving users, we also need to increase
the minimum number of ratings.

{
\renewcommand{\arraystretch}{1.4}
\begin{table}[htb]
\centering {\renewcommand{\tabcolsep}{0.12cm}
{\small
\begin{tabular}{|c|c|c|c|c|c|}
\hline
$m$ & ${\bm \alpha}_i=(\alpha_{i,1},\alpha_{i,2},\alpha_{i,3},\alpha_{i,4},
\alpha_{i,5})$ & $ 1- \delta $ & $\ell'_i$ & $f'_i$ & $n'_i$
\\ \hline \hline
5 & $(\frac{4}{35}, \frac{25}{35}, \frac{3}{35}, \frac{2}{35}, \frac{1}{35})$  & 0.8 & 5 & 0 & 77  \\  \hline
5 & $(\frac{4}{35}, \frac{25}{35}, \frac{3}{35}, \frac{2}{35}, \frac{1}{35})$  & 0.8 & 5 & 0.1 & 93  \\  \hline
5 & $(\frac{4}{35}, \frac{25}{35}, \frac{3}{35}, \frac{2}{35}, \frac{1}{35})$  & 0.8 & 5 & 0.2 & 182  \\  \hline
\end{tabular}
}
} \caption{Biased misbehavior: minimum number of ratings to resist biased misbehavior.}
\label{table:examp_bias_misb}
\end{table}
}

\subsection{Average scoring rule}

{\em Average scoring rule} evaluates the quality of a product by taking
average on {\em all} ratings of that product.  There are number of interesting question
to explore, i.e., how many ratings do we need so that we can produce an average
rating that reflects the quality of a product accurately? What is the effect
of misbehavior in ratings? Let us explore these questions here.

\noindent {\bf Analysis for Honest Rating:} We assume that for the honest rating case,
with a large enough number of ratings, we can produce an average rating that reflects
the quality of a product accurately. Recall that the average rating of product
$P_i$ is $\widehat{r}_i \!=\! \sum_j r_{i,j} / n_i \!=\! \sum_j r^+_{i,j} / n_i$,
and $\gamma_i$ is the ground truth value that reflects the true quality of
product $P_i$.  Intuitively, $\gamma_i$ can be computed by
$\gamma_i \!=\! \lim_{n_i \rightarrow \infty} \widehat{r}_i$.  From
Lemma \ref{lemma:rating_dis}, we have that
$\MP [ r^+_{i,j} \!=\! k ] \!=\! \alpha_{i,k}$, where $k \!=\! 1, \ldots, m$
and $j \!=\! 1, \ldots, n_i$. Thus, we have $\gamma_i \!=\! \sum_k k \alpha_{i,k} $.
Let us first explore the case that all users rate honestly.  We
present the main theorem for the honest rating case as follows.

\begin{theorem}
[Honest Rating Case]{\em Suppose all users rate honestly.  If
product $P_i$ has at least
\begin{equation}
   n_i \geq  n'_i = \frac{3}{\epsilon^2} \ln\frac{2m}{\delta}
\label{ineq:asr_sincere:ni}
\end{equation}
ratings, then
\[
    \left| \widehat{r}_i - \gamma_i \right|
    \leq
    \epsilon \sqrt{ m \gamma_i } + m \epsilon^2
\]
holds with probability at least $1 - \delta$.
\label{theorem:as_sincere} }
\end{theorem}

\noindent {\bf Proof:} Please refer to the appendix for derivation. \done

\noindent{\bf Remark:}  The physical meaning of
Theorem \ref{theorem:as_sincere} is that
with a large enough number of ratings, we can produce an accurate
estimation of $\gamma_i$ with an arbitrarily small error .

Table \ref{table:examp_as} shows some numerical results on the minimum number of
ratings we need to produce an accurate estimation of $\gamma_i$ when all users rate
honestly. Table \ref{table:examp_as} depicts $m$, $\bm \alpha_i$, $1-\delta$, $n'_i$
and the absolute error bound between average rating and its ground truth value,
or $E_r \!=\! \epsilon \sqrt{ m \gamma_i } + m \epsilon^2$ respectively. We could
see that increasing the minimum number of ratings will decrease the error.  It is
also interesting to observe that using the average score rule, we need at least 221
ratings, while one only needs 77 ratings using the majority rule
(please refer to Table \ref{table:mr_Sincere}).

{
\renewcommand{\arraystretch}{1.4}
\begin{table}[htb]
\centering {\renewcommand{\tabcolsep}{0.12cm}
{\small
\begin{tabular}{|c|c|c|c|c|c|}
\hline
$ m $ & ${\bm \alpha}_i=(\alpha_{i,1},\alpha_{i,2},\alpha_{i,3},\alpha_{i,4},
\alpha_{i,5})$ & $ 1- \delta $ & $E_r$ & $n'_i$
\\ \hline \hline
5 & $(\frac{4}{35}, \frac{25}{35}, \frac{3}{35}, \frac{2}{35}, \frac{1}{35})$ & 0.8 & 1 & 221  \\  \hline
5 & $(\frac{4}{35}, \frac{25}{35}, \frac{3}{35}, \frac{2}{35}, \frac{1}{35})$ & 0.8 & 0.75 & 366  \\  \hline
5 & $(\frac{4}{35}, \frac{25}{35}, \frac{3}{35}, \frac{2}{35}, \frac{1}{35})$ & 0.8 & 0.5 & 716  \\  \hline
\end{tabular}
}
} \caption{Honest rating: minimum number of ratings ({\em average
scoring rule}).} \label{table:examp_as}
\end{table}
}

\noindent {\bf Analysis of Rating under Misbehavior:}  We consider
random misbehavior and biased misbehavior specified in last
sub-section.  We treat the ratings of misbehaving users as noise.
The collective rating behavior of the honest users to product $P_i$
is specified by our model with parameter $\bm \alpha_i$ and
$\gamma_i \!=\! \sum_k k \alpha_{i,k}$ is the ground truth value of
average rating of product $P_i$. The following theorems state the
effect of random misbehavior and biased misbehavior respectively.

\begin{theorem}
[Random Misbehavior] {\em  Let $f_i$ denote the fraction of random
misbehaving users who rate product $P_i$. If product $P_i$ has $n_i
\geq 3 \ln(2m/\delta) / \epsilon^2$ ratings, then
\begin{align*}
& \left| m/2 \!-\! \gamma_i \right|\!f_i \!-\!
 \epsilon \sqrt{ m \! \left( \gamma_i \!+\! mf_i/2 - \gamma_i f_i \right)}
 \!-\! m \epsilon^2 \!\leq\! \left| \widehat{r}_i \!-\! \gamma_i \right| \\
&\hspace{0.3 in}
  \leq \epsilon \sqrt{ m \left( \gamma_i \!+\!
    mf_i / 2 - \gamma_i f_i \right) } + m \epsilon^2
    + \left| m/2 \!-\! \gamma_i \right|f_i
\end{align*}
holds with probability at least $1 - \delta$.
\label{theorem:as_random_misb} }
\end{theorem}
\noindent{\bf Proof:} Please refer to the appendix for derivation. \done
\begin{theorem}
[Biased Misbehavior] {\em Let $\ell'_i$ denote the rating that the
biased misbehaving users bias toward.  Let $f'_i $ denote the
fraction of biased misbehaving users who rate product $P_i$. If
product $P_i$ has $n_i \geq 3 \ln(2m/\delta) / \epsilon^2$ ratings,
then
\begin{align*}
& \left|\ell'_i \!-\! \gamma_i \right| f'_i  \!-\!
  \epsilon \sqrt{ m \left( \gamma_i \!+\! f'_i \ell'_i \!-\! \gamma_i f'_i \right) } -
m \epsilon^2 \leq \left| \widehat{r}_i - \gamma_i \right| \\
&\hspace{0.5 in}
  \leq \epsilon \sqrt{ m \left(\gamma_i
     + f'_i \ell'_i - \gamma_i f'_i \right) } + m \epsilon^2
    + \left| \ell'_i - \gamma_i \right| f'_i
\end{align*}
holds with probability at least $1 - \delta$.
\label{theorem:as_bias_misb} }
\end{theorem}
\noindent {\bf Proof:} Please refer to the appendix for derivation.  \done

\noindent {\bf Remark:} The physical meaning of
Theorem \ref{theorem:as_random_misb} and
\ref{theorem:as_bias_misb} is that average scoring rule is {\em
sensitive} to random and biased misbehavior.  If the fraction of
misbehaving users is small, then we can still reflect the true
quality of products with a high probability, otherwise it is
{\em impossible} to accurately reveal the ground truth.

\section{ Inferring Model Parameters  } \label{section:infer_algo}

In last section, we presented the formal analysis of our model with
the assumption that the model's parameters $\bm \alpha_i, \forall i$ are known.
In this section, we show how to obtain these parameters' values
by presenting a maximum likelihood inference
algorithm to {\em infer} these parameters $\bm \alpha_i, \forall i$
from partial information, or historical ratings.
With this inference algorithm, we can apply our framework to analyze and improve the
applications of online rating systems in web services, i.e., eBay, TripAdvisor, etc.

Recall that $\Sr ^{+}_i \!=\! \{r^+_{i,1}, \ldots, r^+_{i, n_i}\}$
represents a set of all observed ratings of product $P_i$.  We
seek to infer $\bm \alpha_i$ from $\Sr^+_i$.  From Lemma
\ref{lemma:rating_dis}, we could write down the probability mass
function (pmf) of $r^+_{i,j}$ as follows: $\MP [ r^+_{i,j} \!=\! k ]
\!=\! \alpha_{i,k}$, where $j \!=\! 1, \ldots, n_i$ and $k \!=\! 1,
\ldots, m$.  Let $n_{i,k}$ denote the number of ratings of product
$P_i$ that equal to $k$. Then the likelihood of
the parameter ${\bm \alpha}_i$ given a set of observed ratings $\Sr
^{+}_i$ can be expressed as:
\begin{align*}
    \mathcal{L}({\bm \alpha}_i)
    & = p( \Sr^+_i | {\bm \alpha}_i )
    = \prod\nolimits_{j=1}^{n_i} p(r^+_{i,j} | {\bm \alpha}_i )
    = \prod\nolimits_{k=1}^m (\alpha_{i,k})^{n_{i,k}}.
\end{align*}
The remaining issue is to derive the maximum likelihood estimation
of the parameter ${\bm \alpha}_i$, which is denoted by $\widehat{\bm
\alpha}_i$, via maximizing $\mathcal{L}({\bm \alpha}_i)$. This is
equivalent to maximize the log likelihood function:
\begin{align*}
    L({\bm \alpha}_i) & = \log \mathcal{L}({\bm \alpha}_i)
                       = \sum\nolimits_{k=1}^m n_{i,k} \log \alpha_{i,k} \\
                      & = \sum\nolimits_{k=1}^{m-1} n_{i,k} \log \alpha_{i,k} +
                           n_{i,m} \log (1-\sum\nolimits_{k=1}^{m-1}\alpha_{i,k} ).
\end{align*}
By maximizing $L({\bm \alpha}_i)$, we obtain the maximum likelihood
estimation of $\bm \alpha_i$ as follows:
\[
    \widehat{\alpha}_{i,k} = \frac{n_{i,k}}{n_i},
    \hspace{0.1 in}
    \mbox{for $k = 1, \ldots, m$ }.
\]
The inference algorithm is outlined in Algorithm \ref{infer_paprameter}.
\begin{algorithm}
\caption{Inference algorithm for model parameters}
\label{infer_paprameter}
\begin{algorithmic}[1]
    \REQUIRE A set of ratings of product $P_i$: $\Sr^+_i =\{r^+_{i,1}, \ldots, r^+_{i,n_i}\}$
    \ENSURE  $\widehat{{\bm \alpha}}_i$
    \STATE $n_{i,k} = |\{r^+_{i,j} | r^+_{i,j} \in \Sr^+_i, r^+_{i,j} = k \}|$, $k = 1,\ldots, m$
    \FOR{ $k$ = 1 to $m$}
        \STATE $\widehat{\alpha}_{i,k} = n_{i,k} / n_i $
    \ENDFOR
\end{algorithmic}
\end{algorithm}

\noindent {\bf Remark:} The running time of this algorithm is
$\Theta (\sum_{i=1}^N |\Sr^+_i|)$, or the running time is {\em
linear} with respect to the number of ratings.

\section{Experiments on Synthetic Data } \label{section:exp_syn_data}
Let us first use synthetic data to examine various factors that
influence the accuracy of product quality evaluation. This will help
us gain some insights on improving applications of online rating
systems. We consider an online rating system with a 5-level cardinal
rating metric, say $m \!=\! 5$. We start our evaluation from the
context where the {\em majority rule} is used to aggregate ratings.
Then we discuss the context corresponding to the {\em average
scoring rule}.

\subsection{  Majority rule}

We start with the case that all users are honest, then explore when
some users misbehave in their ratings.

\begin{figure}[htb]
\centering
\begin{minipage}[t]{0.48\linewidth}
\includegraphics[width=\textwidth]{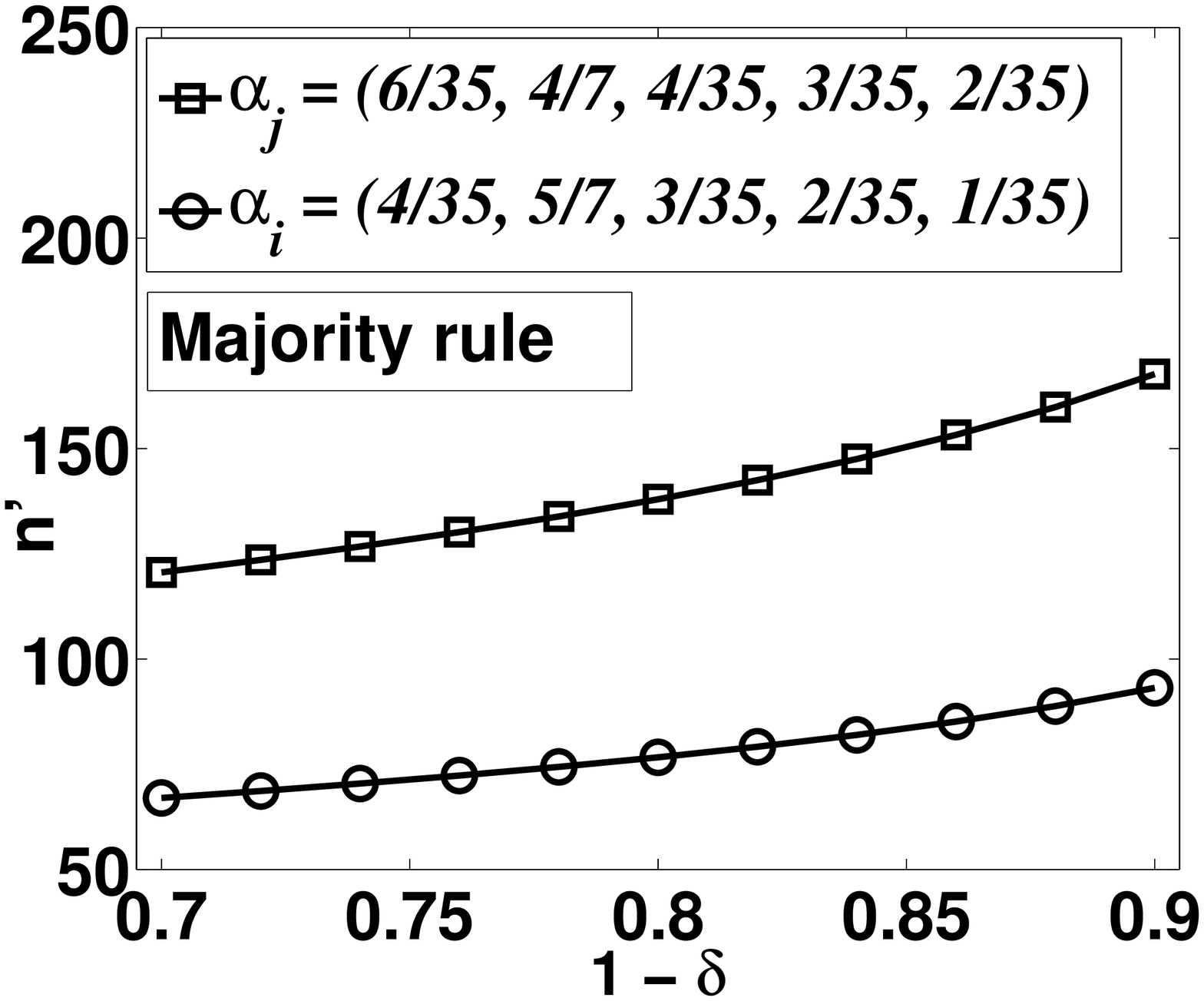}
\caption{Impact of model parameter ${\bm \alpha}_i$ and success probability $1-\delta$.}
\label{fig:mr_sincere}
\end{minipage}
\hspace{0.05in}
\begin{minipage}[t]{0.48\linewidth}
\centering
\includegraphics[width=\textwidth]{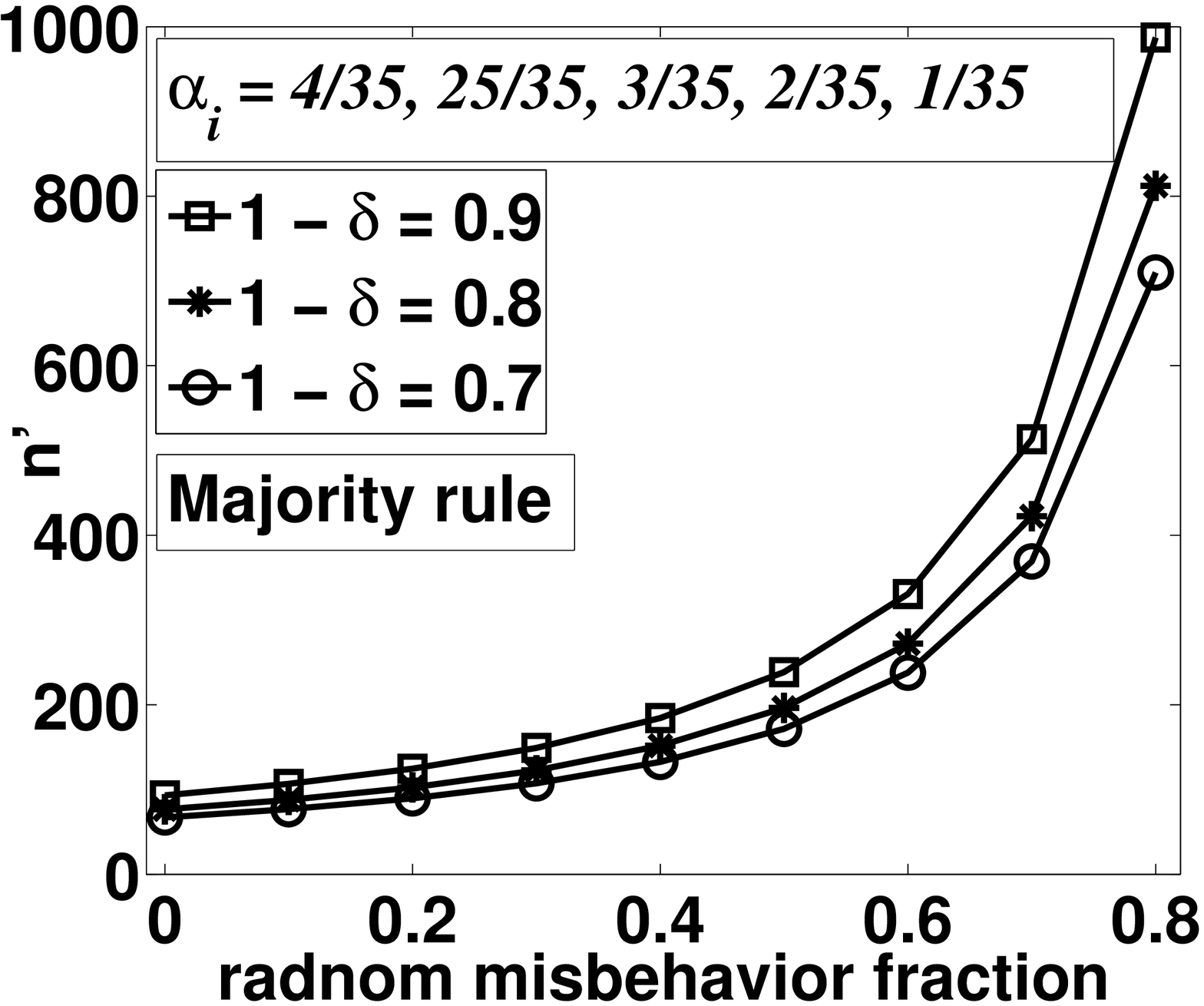}
\caption{Impact of random misbehavior.
}
\label{fig:mr_random_misb}
\end{minipage}
\end{figure}

\noindent {\bf Experiment 1: Effect of Model Parameter $\bm
\alpha_i$ and Success Probability.}  In this experiment, we study
the honest ratings case where all users rate honestly.  To examine
the effect of model parameters, we synthesize the ratings of two
products, say $P_i$ and $P_j$, using our model with parameter ${\bm
\alpha}_i \!=\! (\frac{4}{35}, \frac{5}{7}, \frac{3}{35}, \frac{2}{35}, \frac{1}{35})$
and ${\bm \alpha}_j \!=\!(\frac{6}{35}, \frac{4}{7}, \frac{4}{35},
\frac{3}{35}, \frac{2}{35})$ respectively. Let $n'$ denote the minimum number of
ratings we need to extract the true label of a product with success
probability of at least $1 - \delta$. To examine the effect of the
success probability, we vary $1-\delta$ from 0.7 to 0.9. The
numerical results of $n'$ are shown in Fig.~\ref{fig:mr_sincere},
where the horizontal axis represents the success probability, or $1
\!-\! \delta$, and the vertical axis shows the corresponding minimum
number of ratings, or $n'$. From Fig.~\ref{fig:mr_sincere} we have
the following observations. First, when we increase the success
probability, or $1 - \delta$, the minimum number of ratings
increased.  In other words, to improve the success probability, we
need a larger the number of ratings.
Note that the minimum number of ratings increases only with a {\em
small} rate. To extract the true label with the same probability, we
need far more ratings for product $P_j$ (with a higher variance
in rating scores than $P_i$) than $P_i$.

\noindent {\bf Lessons learned:}  Increasing the success probability
increases the minimum number of ratings.  The
variance on rating scores of a product has a
significant impact on the minimum number of ratings. So it is
interesting to explore further, i.e., are these results sensitive to
misbehavior? Let us continue to explore.

\noindent {\bf Experiment 2: Effect of Random Misbehavior.}  We
extend the online rating system specified in Exp.~1 to a system
where the user population consists of honest users
and random misbehaving users.
We synthesize the ratings of product $P_i$ from the
honest users using our model with parameter ${\bm \alpha}_i \!=\!
(\frac{4}{35}, \frac{5}{7}, \frac{3}{35}, \frac{2}{35}, \frac{1}{35})$.  Here $n'$ represents the same measure as specified
in Exp.~1.  To examine the effect of random misbehavior, we vary the
fraction of random misbehaving users from 0 to 0.8. The numerical
results of $n'$ are shown in Fig.~\ref{fig:mr_random_misb}
in which three curves correspond to success probability of at
least 0.7, 0.8, 0.9 respectively. We can observe that when the fraction of misbehavior
is less than 0.6, increasing the fraction of random misbehaving users only
increases the minimum number of ratings slightly.  When
the fraction of misbehavior is larger than 0.6, a small increase in
the fraction of random misbehavior can increase the minimum number of ratings remarkably.

\noindent {\bf Lessons learned:} When random misbehaving users exist,
one can still extract the true label of a product with a price of increasing
the minimum number of ratings.  A small fraction of random misbehavior only
increases the minimum number of ratings slightly.
This shows the robustness of {\em majority rule}.  But
when this fraction is increased beyond a certain threshold, we need
a lot more ratings to compensate.
Next, we explore the impact of biased misbehavior in users' ratings.

\begin{figure}[htb]
\centering
\begin{minipage}[t]{0.48\linewidth}
\includegraphics[width=\textwidth]{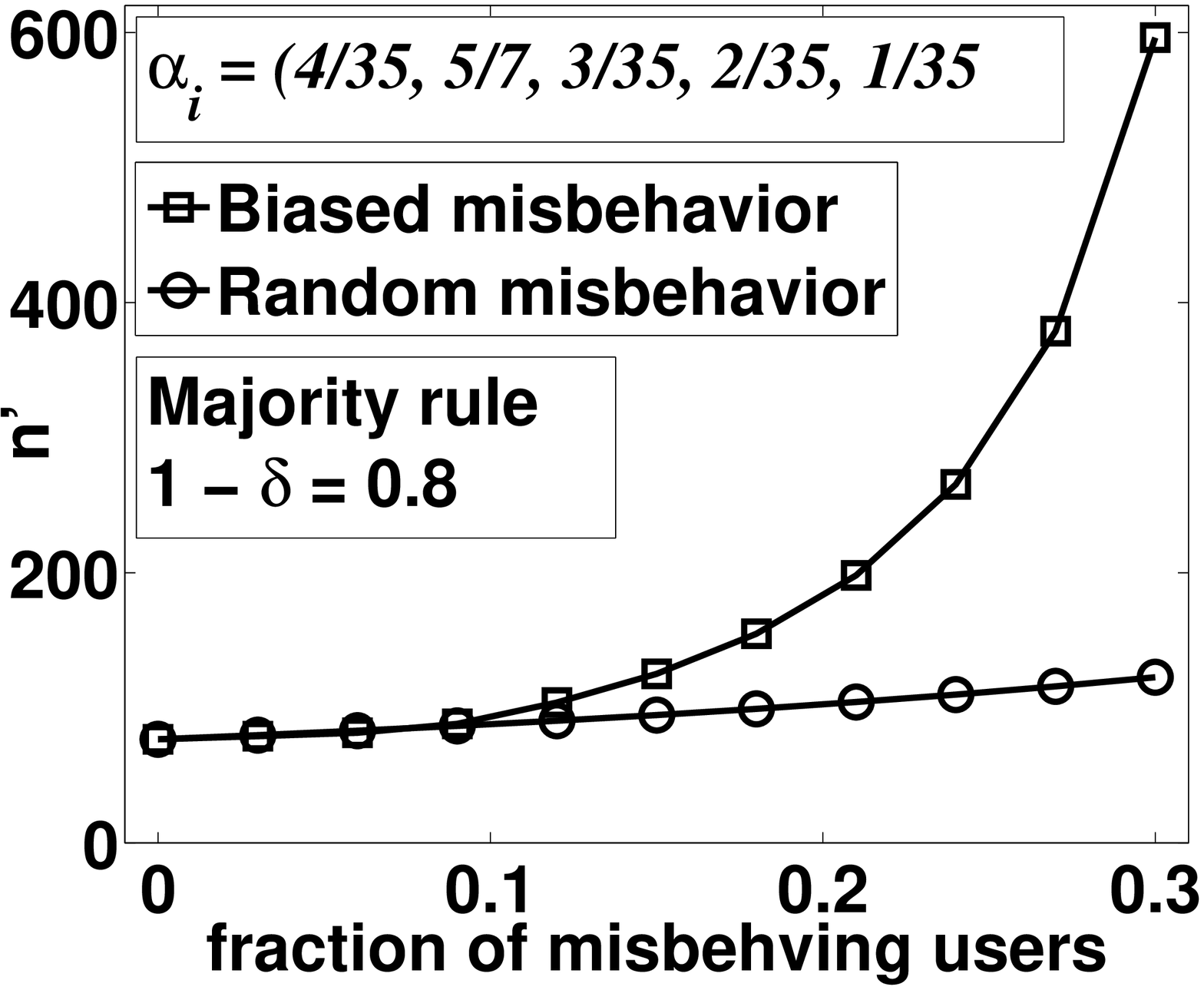}
\caption{Impact of biased misbehavior.
}
\label{fig:mr_bias_misb_less}
\end{minipage}
\hspace{0.05in}
\begin{minipage}[t]{0.48\linewidth}
\centering
\includegraphics[width=\textwidth]{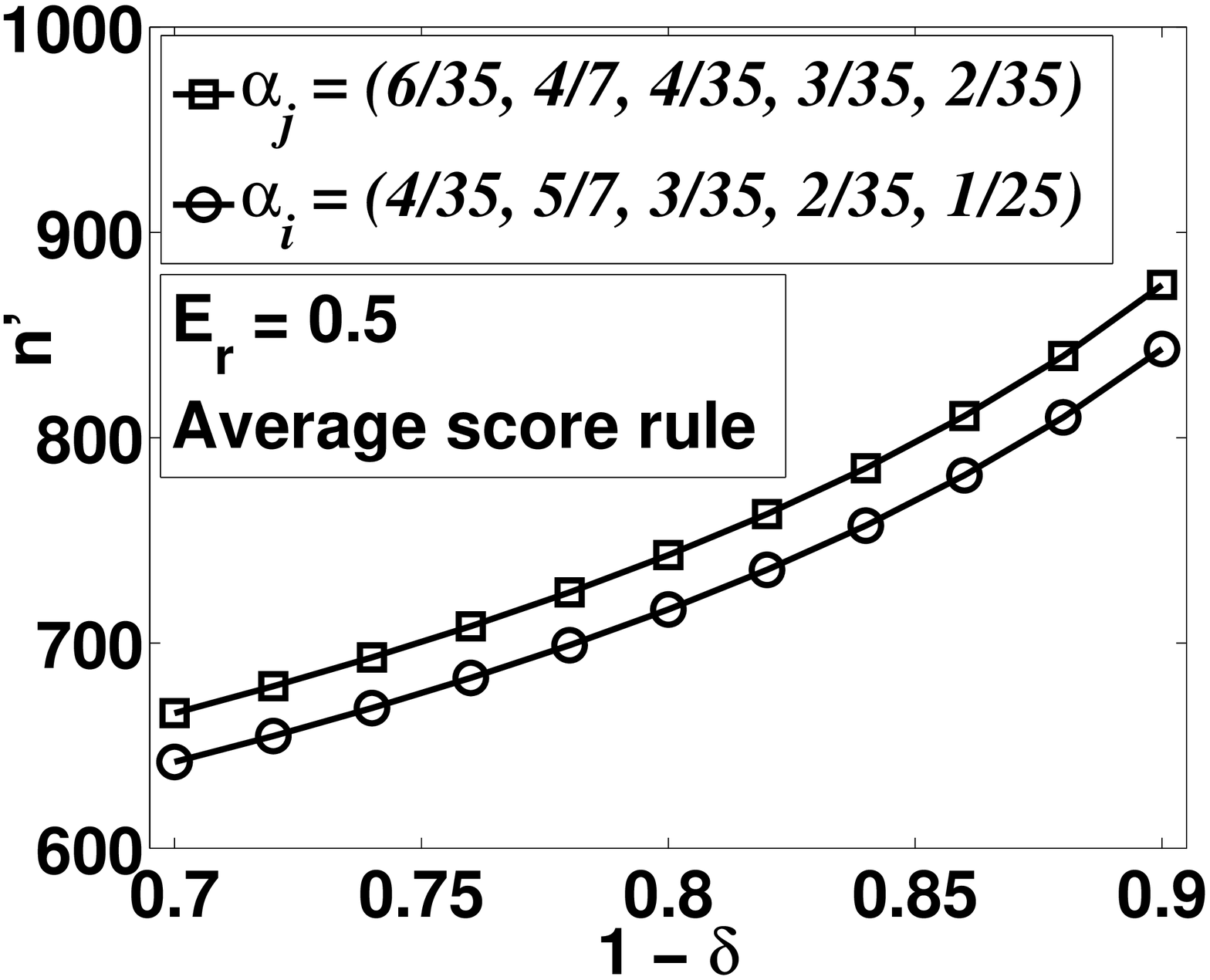}
\caption{Impact of model parameter ${\bm \alpha}_i$ and success probability $1 - \delta$.
}
\label{fig:as_mp}
\end{minipage}
\end{figure}

\noindent {\bf Experiment 3: Effect of Biased Misbehavior.}
We extend the online rating system specified in Exp.~1 to the
system where the user population consists of honest users and biased misbehaving users
who bias toward one specific rating.  We synthesize the ratings of product
$P_i$ from the honest users using our model with
parameter ${\bm \alpha}_i \!=\! (\frac{4}{35}, \frac{5}{7}, \frac{3}{35},
\frac{2}{35}, \frac{1}{35})$. The biased misbehaving users bias toward rating 5.  From
Theorem \ref{theorem:mr_bias:bias_win} and \ref{theorem:mr_bias:extract_true_label},
one can still extract the true label with a high probability if the fraction of
biased misbehaving users is less than 0.407.  But if that fraction is larger than 0.407,
then it is impossible to extract the true label with high probability.
We vary the fraction of biased misbehaving users from 0 to 0.3.
Here $n'$ represents the same measure as specified in Exp.~1.
The numerical results of $n'$ are shown in Fig.~\ref{fig:mr_bias_misb_less}.
To compare the impact of random misbehavior and biased misbehavior, we draw another
curve corresponding to random misbehavior in
Fig.~\ref{fig:mr_bias_misb_less} also.  From Fig.~\ref{fig:mr_bias_misb_less}
we have the following observations. When the fraction of biased misbehaving users
is small, say less than 0.1, increasing the fraction of biased misbehaving users only
increases the minimum number of ratings slightly.  When that fraction is larger
than 0.1, a small increase in the fraction of biased misbehaving users can
remarkably increase the minimum number of ratings needed.
The {\em majority rule} is more robust against random misbehavior compared to
biased misbehavior, since the curve corresponding to random misbehavior is quite flat.

\noindent {\bf Lessons learned:} {\em Majority rule } is robust and
can resist a small fraction, say less than 0.1, of biased misbehaving users.
Also, the majority rule is robust against random misbehavior.
The biased misbehavior is more disruptive
than the random misbehavior in distorting the evaluation process.

\subsection{ Average scoring rule}

We start our study with a simple case, say all users rate honestly.

\noindent {\bf Experiment 4: Effect of Model Parameters $\bm \alpha_i$ and Success
Probability.}  We study the case where all users rate honestly.
To examine the effect of model parameters, we synthesize the ratings of two products,
say $P_i$ and $P_j$, using our model with parameter ${\bm \alpha}_i \!=\!
(\frac{4}{35}, \frac{5}{7}, \frac{3}{35}, \frac{2}{35}, \frac{1}{35})$
and ${\bm \alpha}_j \!=\! (\frac{6}{35}, \frac{4}{7}, \frac{4}{35},
\frac{3}{35}, \frac{2}{35})$ respectively.
Let $E_r$ denote the absolute error bound between the average rating and its ground truth
value.  We set $E_r \!=\! 0.5$.  Let $n'$ denote the minimum number of ratings a
product needs so as to guarantee that the absolute error is less or equal
to $E_r$ with probability of at least $1 \!-\! \delta$.
We vary the success probability $1-\delta$ from 0.7 to 0.9.
The numerical results of $n'$ are shown in Fig.~\ref{fig:as_mp}, where the horizontal
axis represents the success probability, or $1 \!-\! \delta$, and the vertical
axis shows the corresponding minimum number of ratings, or $n'$.
We have the following observations.
The minimum number of ratings is insensitive to the given
parameters $\bm \alpha_i$ and $\bm \alpha_j$, since these two curves in
Fig.~\ref{fig:as_mp} similar.  When we increase the success probability,
we increase the minimum number of ratings slightly.

Now we compare the performance of the two score aggregation rules we study.
We compare the minimum number of ratings for the {\em majority rule} and
the {\em average scoring rule} and the results are shown in
Table \ref{table:compare_to_majority}.  One can see that to achieve the same
level of confidence, we need a lot more ratings for {average scoring rule}.

\noindent {\bf Lessons learned:}  To obtain a reliable evaluation of product quality
using the {\em average scoring rule}, we need remarkably more ratings than using
the {\em majority rule}.  The minimum number of ratings is insensitive to model
parameters or success probability.  This implies that the majority rule requires
less ratings to guarantee the same level of confidence, so it is
the preferred rating aggregation rule.

\begin{table}
\centering
\begin{tabular}{|c||c|c|c|c|c|} \hline
$1 - \delta$ & 0.7 & 0.75 & 0.8 & 0.85 & 0.9 \\  \hline \hline
$n'$ (majority) & 67 & 70 & 77 & 83 & 93  \\  \hline
$n'$ (average score) & 642 & 675 & 716 & 769 & 843    \\  \hline
\end{tabular}
\caption{ Comparison of minimum number of ratings for {\em majority rule} and {\em average scoring rule}.} \label{table:compare_to_majority}
\end{table}

\noindent {\bf Effect of Misbehavior.}  To examine the effect of misbehavior,
we extend the online rating system as specified in Exp.~4: one consists of
honest users and random misbehaving users, and the other one consists of honest users
and biased misbehaving users.  We synthesize the ratings of product $P_i$ from the
honest users using our model with parameter ${\bm \alpha}_i \!=\!
(\frac{4}{35}, \frac{5}{7}, \frac{3}{35}, \frac{2}{35}, \frac{1}{35})$.
The biased misbehaving users bias toward rating 5.  For both extensions, we
vary the fraction of misbehaving users from 0.1 to 0.3.  Here $n'$ and $E_r$
(set to 0.5) represents the same measure as specified in Exp.~4.
The numerical results of $n'$ are shown in Table \ref{table:impact_misb}.
To compare the robustness of the {\em majority rule} and the {\em average scoring rule},
we show the numerical results of $n'$ corresponding to the {\em majority rule} in the
same table.  From Table \ref{table:impact_misb}, we could see that the
{\em majority rule} is more robust against misbehavior (either random misbehavior
or biased misbehavior), since to resist the same fraction of misbehaving users,
the {\em average scoring rule} requires a lot more ratings.

\begin{table}
\centering
\begin{tabular}{|c||c|c|c|c|c|} \hline
\multicolumn{6}{|c|}{ \bf Random misbehavior }  \\ \hline
fraction of bad users& 0.1 & 0.15 & 0.2 & 0.25 & 0.3 \\  \hline \hline
$n'$ (majority) & 89 & 94 & 102 & 112 & 122  \\  \hline
$n'$ (average score) & 842 & 882 & 947 & 1018 & 1099    \\  \hline \hline \hline
\multicolumn{6}{|c|}{ \bf Biased misbehavior }  \\ \hline
fraction of bad users& 0.1 & 0.15 & 0.2 & 0.25 & 0.3  \\  \hline \hline
$n'$ (majority) & 93 & 125 & 181 & 296 & 259    \\  \hline
$n'$ (average score) & 1107 & 1452 & 2027 & 3110 & 5600    \\  \hline
\end{tabular}
\caption{Impact of misbehavior on two rating aggregation rules}
\label{table:impact_misb}
\end{table}

\section{Experiments on real data } \label{section:exp_real_data}

In this section, we show the experimental results on three
{\em large} datasets which we obtain via web crawling:
hotel ratings from TripAdvisor, product ratings from Amazon,
and seller ratings from Ebay. We first validate our model with these three datasets,
and then explore a number of important questions, i.e., what's the minimum number
of ratings a hotel/product/seller needs to receive so
that one can evaluate its quality accurately?

\subsection{ Datasets}


\noindent{\bf TripAdvisor.} It is one of the most popular travel websites
which assists customers in booking hotels, restaurants, etc, and
post travel-related opinions in the form of ratings (or reviews) to hotels, restaurants,
etc.  We crawled the historical ratings of 11,540 hotels.
Table~\ref{data:data_stat} shows the overall statistics of the data set.

\noindent{\bf Amazon.} Amazon is one of the most popular recommendation systems that
assists customers in product adoption, and one can post opinions on products in the
form of ratings (or reviews). We crawled the historical ratings of 32,888 products.
Table \ref{data:data_stat} shows the overall statistics of the data set.

\noindent{\bf Ebay.} Ebay is an E-commerce system that assists customers
in online product purchasing, and customers can post opinions of sellers in
the form of ratings (or reviews) to reflect the reputation of a seller.
We crawled the historical ratings of 4,586 sellers.
Table \ref{data:data_stat} shows the overall statistics of the data set.

{\renewcommand{\arraystretch}{1.3}
\begin{table}[htb]
\centering
{\small
\begin{tabular}{|c||c|c|c|} \hline
& {\bf TripAdvisor} & {\bf Amazon} & {\bf Ebay} \\ \hline
{\bf Number of items } & 11,540  & 32,888 & 4,586 \\ \hline
{\bf Total number of ratings} & 3,114,876 & 5,066,070 & 19,217,083 \\  \hline
{\bf Maximum / Minimum on} & 9930/1 & 24,195/1 &  117,100/1 \\
{\bf number of ratings }       &        &          &            \\ \hline
{\bf Mean / Median on}     & 269.9/179  & 154/47 & 4190/1437 \\
{\bf number  of ratings }       &        &          &            \\ \hline
{\bf Rating metricr:} $\{1,...,m\}$& $1, \ldots, 5$ & $1, \ldots, 5$ & $1, \ldots, 12$ \\
\hline\end{tabular}
}
\caption{Statistics for three rating data sets.} \label{data:data_stat}
\end{table}
}

\subsection{ Model validation}
\label{section:model_vald}
We validate our model by showing that if an item
meets the requirement on minimum number of ratings,
then the evaluation of its quality is indeed reliable.

\noindent {\bf Extract minimum number of ratings.}
To extract the minimum number of ratings, we need to first estimate model
parameters ${\bm \alpha_i}, \forall i$.  Since the accuracy depends on the number of
ratings, therefore, we select those items with a large number of ratings, i.e., those
with at least 400 ratings.  Table~\ref{data:data_selected_item} shows the number of
items in the dataset satisfy this selection criteria.  From
Table~\ref{data:data_selected_item} we could see that in total we select
$2368+2396+3307 \!=\! 8071$ items out.  We map the ID of these 8071 items to
$1,..., 8071$ and use $P_1, \ldots, P_{8071}$ to denote them.  For each selected item
$P_i, i \in \{1, \ldots, 8071\}$, we applly Algorithm~\ref{infer_paprameter} on their
rating sets, so as to obtain an estimation of its model parameter denoted by
$\widehat{{\bm \alpha}}_i$.  We use the inferred parameters $\widehat{{\bm \alpha}}_i$
to compute the minimum number of ratings for item $P_i, i \in \{1, \ldots, 8071\}$,
setting the success probability $1\! -\! \delta \!=\! 0.8$, and $E_r = 0.5$.
We use $n'$ to denote the desired minimum number of ratings an item needs.

\noindent {\bf Model validation algorithm design.} The dataset is with time stamps.
Based on this, our validation procedure can be stated as follows.  For each time stamp
we first check if the number of ratings of a hotel (up to that time stamp) meets the
requirement on minimum number of ratings.  If yes, we then apply a rating aggregation
rule to evaluate its quality (based on the historical ratings up to that time stamp),
and check if the evaluation is reliable.  After we finished the whole process,
we compute the faction of evaluations which are reliable.  We outline our model
validation algorithm in Algorithm~\ref{algo:model_valid}.  We use the following
notations in the algorithm.  For each hotel, we sort its ratings based on ratings'
times tamps.  Let $\tau_i \!=\! \{\tau_{i,1}, \ldots, \tau_{i, n_i}\}$ denote a set of
sorted ratings of hotel $P_i$, where $i \!=\! 1, \ldots, 8071$, $n_i$ is the number of
ratings of $P_i$, and $\tau_{i,1}$ is the earliest rating.  Let $N_{test}$ denote
the number of time stamps that an item meets the desired minimum number of ratings,
$N_{reliable}$ denote the number of reliable evaluations, and the faction of reliable
evaluations is denoted by $f_{reliable}$.

{\renewcommand{\arraystretch}{1.3}
\begin{table}[htb]
\centering
{\small
\begin{tabular}{|c||c|c|c|} \hline
& {\bf TripAdvisor} & {\bf Amazon} & {\bf Ebay} \\ \hline
{\bf $\#$ of selected items } & 2368  & 2396 & 3307 \\ \hline
\end{tabular}
}
\caption{Number of selected items.} \label{data:data_selected_item}
\end{table}
}

We apply Algorithm~\ref{algo:model_valid} on TripAdvisor, Amazon and Ebay datasets
respectively, and we compute the faction of reliable evaluations $f_{reliable}$ for
each of them seperately.  The results of $f_{reliable}$ are shown in
Table \ref{table:model_valid}, which depicts rating aggregation rules, the number of
time stamps a hotel meets the requirement of minimum number of ratings $N_{test}$,
the number of reliable evaluations $N_{reliable}$, and the faction of reliable
evaluations $f_{reliable}$.  From Table \ref{table:model_valid}, we have two
observations.  First, if a product meets the requirement on minimum number of ratings,
then the evaluation of its quality is indeed reliable, because the value of
$f_{reliable}$ is around 99\%.  This shows the correctness of our model and our
mathematical framework.  Secondly, it shows the {\em robustness} of the majority rule
since we have more time stamps (or ratings) that satisfy the minimum requirement.
In fact, the {\em majority rule} has around three times
(836088 / 296309 $\approx$ 2.82, TripAdvisor),
two times (2246804 / 981886 $\approx$ 2.29, Amazon), and
1.28 times (16667935 / 13008919 $\approx$ 1.28, Ebay) the number of ratings that can
be used for reliable evaluation.  We explore why this is the case in the next subsection.
\begin{algorithm}
\caption{Algorithm for model validation}
\label{algo:model_valid}
\begin{algorithmic}[1]
    \REQUIRE $\tau_{i_1}, \ldots, \tau_{i_n}$, $\widehat{{\bm \alpha}}_{i_1}, \ldots, \widehat{{\bm \alpha}}_{i_n}$, rating aggregation rule $\SA$
    \ENSURE  $N_{test}, N_{reliable}, f_{reliable}$
    \STATE $N_{test} \leftarrow 0$, $N_{reliable} \leftarrow 0$
    \FOR{ $\kappa \in \{i_1, \ldots, i_n\}$}
        \STATE Extract the true quality of $P_i$, say if $\SA$ is majority rule, then the true quality is $\gamma_i = \arg\max_k\{\widehat{{\alpha}}_{i_1,k}\} $, and if $\SA$ is average score rule, then the true quality is $\gamma_i = \sum_k k \widehat{{\alpha}}_{i_1,k}$
        \STATE $n' \leftarrow \mbox{minimum number of ratings $P_i$ needs}$
        \FOR{ $j$ = $n'$ to $|\tau_\kappa|$ }
            \STATE $N_{test} \leftarrow N_{test} + 1$
            \IF {$\SA (\{\tau_{i,1}, \ldots, \tau_{i,j}\})$ reflects the true quality of $P_i$}
                \STATE $N_{reliable} \leftarrow N_{reliable} + 1$
            \ENDIF
        \ENDFOR
    \ENDFOR
    \STATE $f_{reliable} = N_{reliable} / N_{test}$
\end{algorithmic}
\end{algorithm}

\begin{table}[htb]
\centering
\begin{tabular}{|c||c|c|c|} \hline
{\bf Aggregation Rule} & $N_{test}$ & $N_{reliable}$ & $f_{reliable}$  \\  \hline \hline
\multicolumn{4}{|c|}{ \bf TripAdvisor }  \\ \hline
{\bf Majority rule} & 836088 & 835822 & 99.97\%   \\  \hline
{\bf Average scoring rule}  & 296309 & 295854  & 99.85\%  \\  \hline \hline
\multicolumn{4}{|c|}{ \bf Amazon }  \\ \hline
{\bf Majority rule} & 2246804 & 2241196 & 99.75\%   \\  \hline
{\bf Average scoring rule}  & 981886 & 971987  & 98.99\%  \\  \hline \hline
\multicolumn{4}{|c|}{ \bf Ebay }  \\ \hline
{\bf Majority rule} & 16705440 & 16667935 & 99.78\%   \\  \hline
{\bf Average scoring rule}  & 13008919 & 12990199  & 99.86\%  \\  \hline
\end{tabular}
\caption{Fraction of reliable evaluations when items meet the
requirement on minimum number of ratings.} \label{table:model_valid}
\end{table}

\subsection{Examine minimum number of ratings}
\label{section:examine_min_num_rat}

We consider the same set of items and we use the same settings specified in the
previous sub-section.  Again, $n'$ denotes the required minimum number of ratings
a product (hotel) needs.  The statistics of $n'$ across items are shown in
Table~\ref{table:min_num_rat}, where $f_{[0,400]}$ denotes the fraction of items
with $n' \!\in\! [0,400]$, and we use {\bf MR} and {\bf ASR} to
denote {\em majority rule} and {\em average score rule} respectively
for brevity.  From Table~\ref{table:min_num_rat}, we have the following observations.
For the {\em average scoring rule}, nearly all items need more than 800 ratings.
In fact 99.45\% hotels (TripAdvisor), 99.08\% products (Amazon),
and 99.97\% sellers (Ebay) need more than 800 rating.  And a large fraction of them
need more than 1200 ratings, say 47.55\% hotels (TripAdvisor), 61.56\% products (Amazon)
and 98.34\% sellers (Ebay).  But for the {\em majority rule}, more than half the hotels
only need less than 800 ratings, say 58.79\% hotels, 88.22\% products, 82.8\% sellers
need less than 800 ratings.  What's more, a large fraction of them need less than 400
ratings, i.e., around 39.74\% hotels, 80.51\% products and 58.06\% sellers.
In other words, the minimum number of ratings trend to be of smaller values for
the {\em majority rule}.

\begin{figure}[htb]
\centering
\begin{minipage}[t]{0.45\linewidth}
\includegraphics[width=\textwidth]{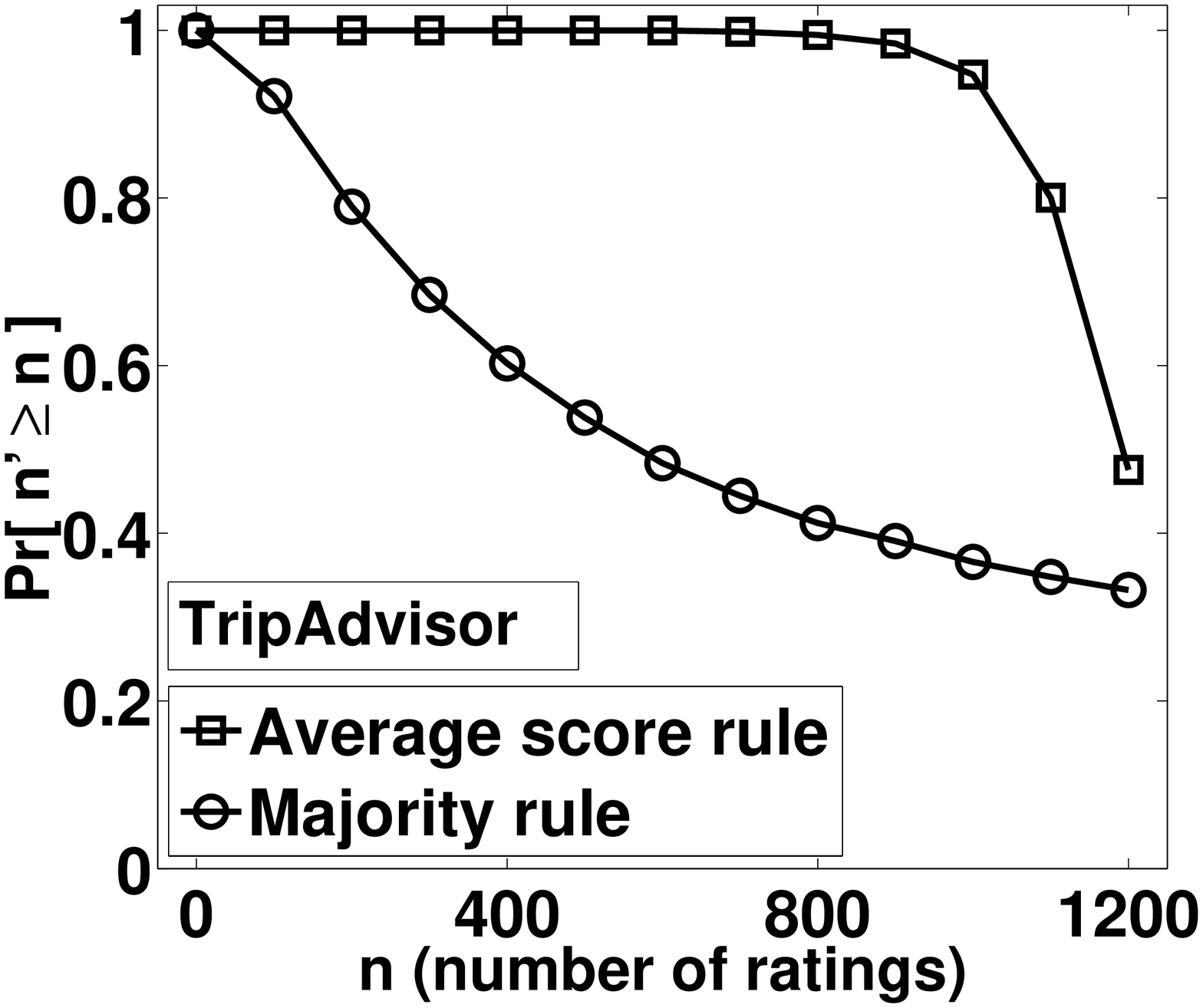}
\caption{Distribution of minimum number of ratings (TripAdvisor)}
\label{fig:eval_rule_tripadvisor}
\end{minipage}
\hspace{0.05 in}
\begin{minipage}[t]{0.45\linewidth}
\centering
\includegraphics[width=\textwidth]{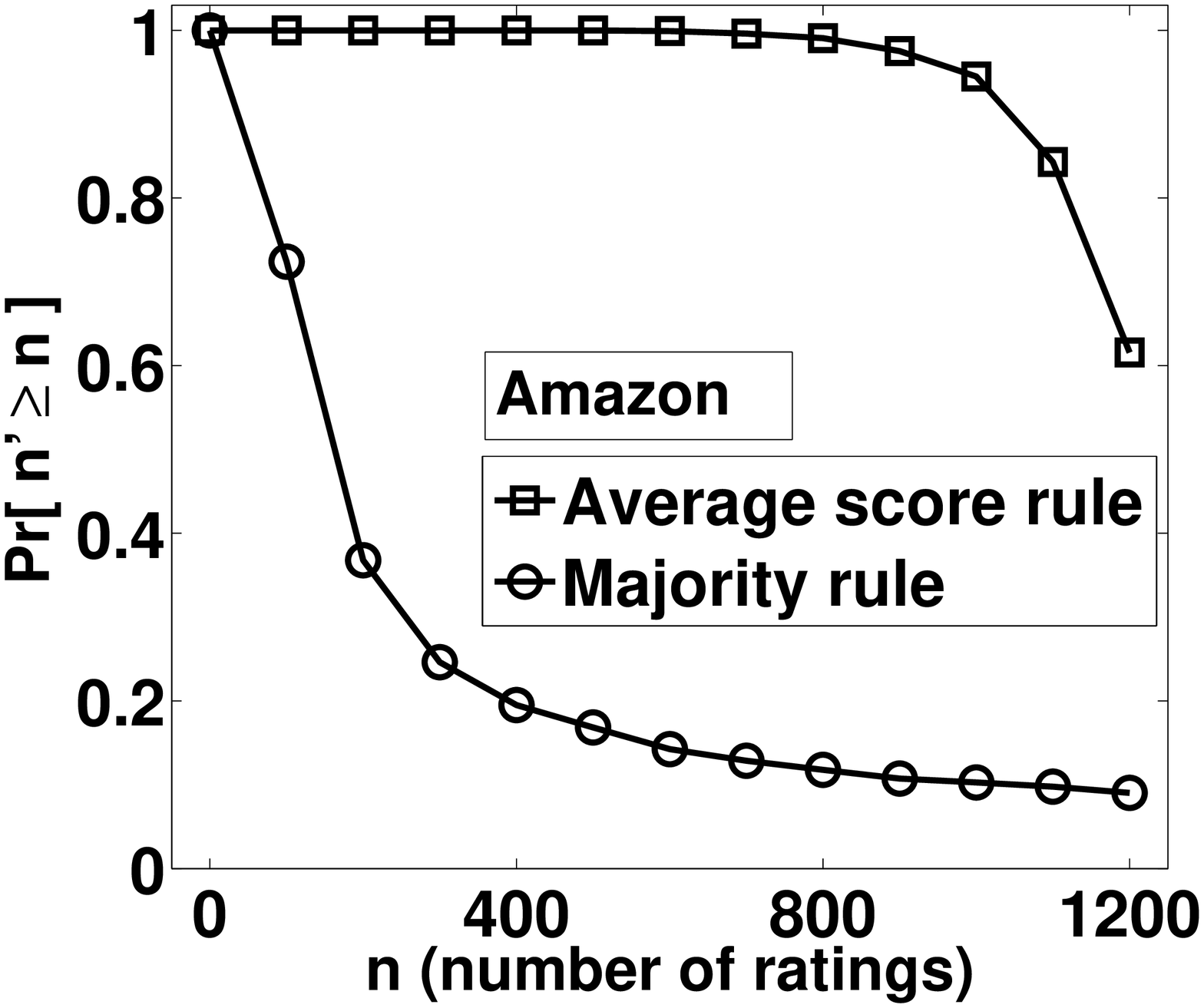}
\caption{Distribution of minimum number of ratings (Amazon)}
\label{fig:eval_rule_amazon}
\end{minipage}
\end{figure}

\begin{figure}[htb]
\centering
\begin{minipage}[t]{0.45\linewidth}
\includegraphics[width=\textwidth]{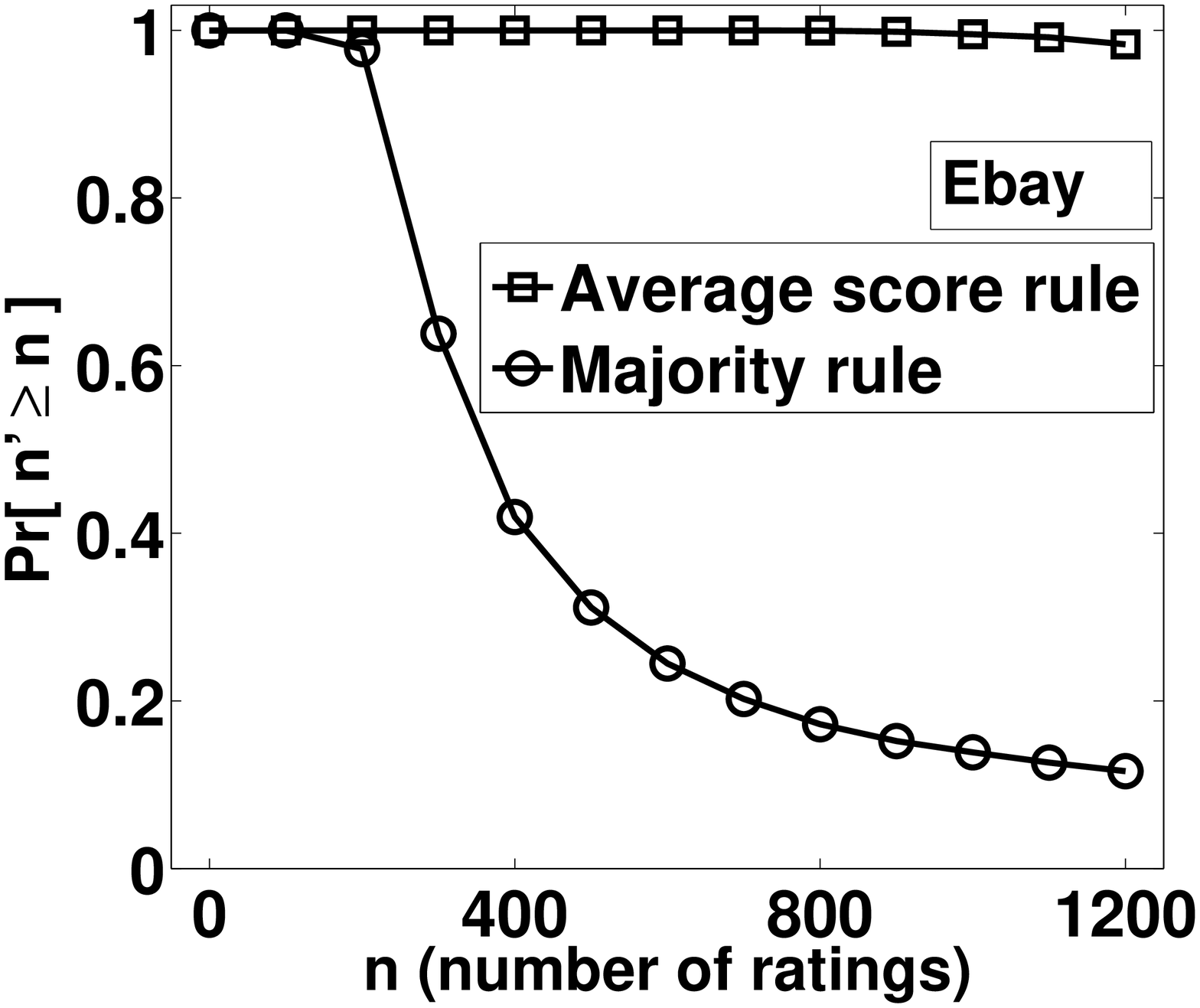}
\caption{Distribution of minimum number of ratings (Ebay)}
\label{fig:eval_rule_ebay}
\end{minipage}
\hspace{0.05 in}
\begin{minipage}[t]{0.45\linewidth}
\centering
\includegraphics[width=\textwidth]{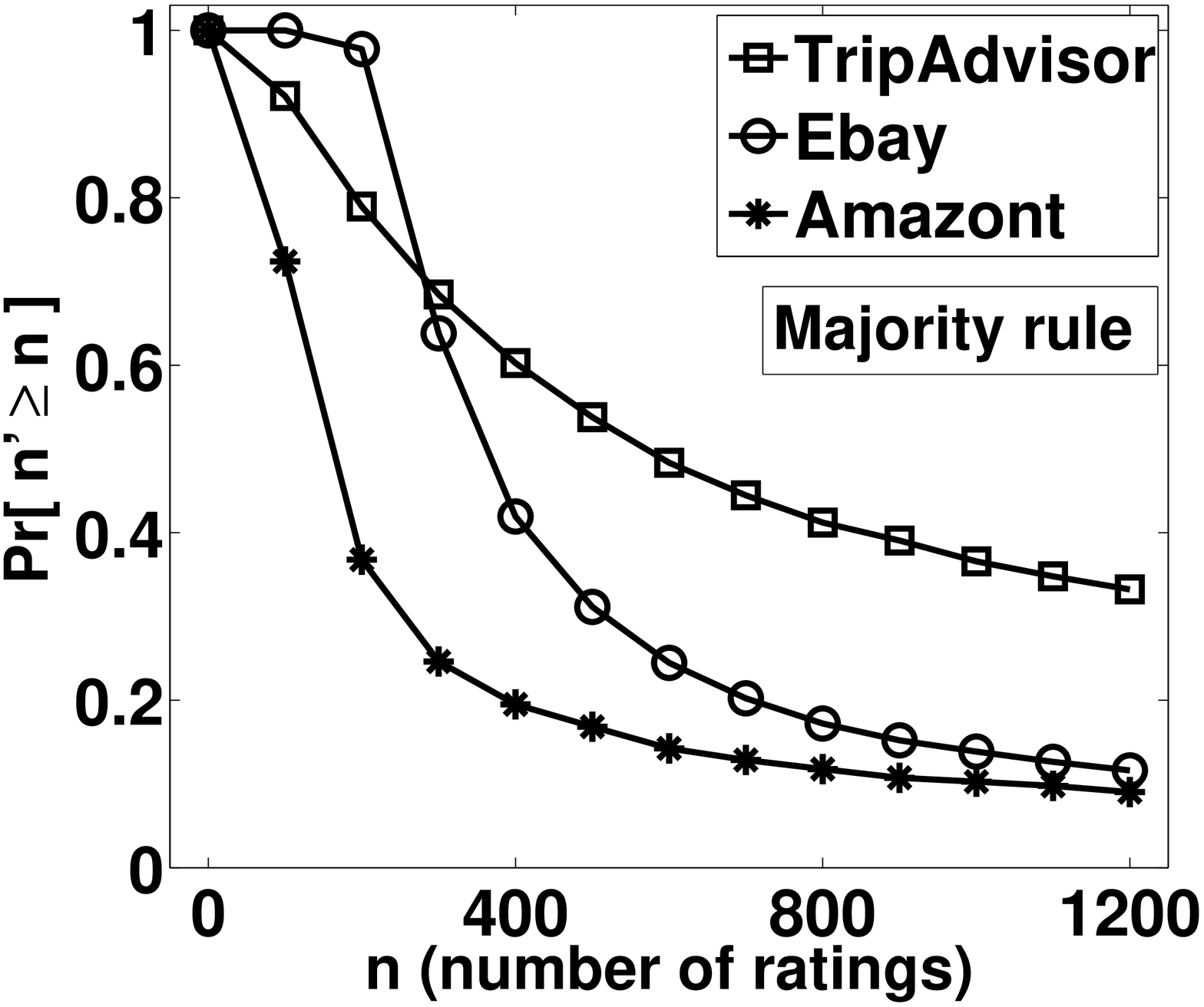}
\caption{Comparing reliability of TripAdvisor, Amazon and Ebay}
\label{fig:reliability_websites}
\end{minipage}
\end{figure}

\noindent {\bf Evaluating rating aggregation rules.} We compare robustness of
{\em majority rule} and {\em average score rule} by examining the distribution of
minimum number of ratings corresponding to them.  We use notation
$\mbox{Pr}[n' \!\geq\! n]$ to denote the fraction of items with minimum number of
ratings larger or equal to $n$.  Fig.~\ref{fig:eval_rule_tripadvisor},
\ref{fig:eval_rule_amazon} and \ref{fig:eval_rule_ebay} show the numerical results of
$\mbox{Pr}[n' \!\geq\! n]$ corresponding to TripAdvisor, Amazon and Ebay dataset
respectively, where the horizontal axis shows the number of ratings $n$, and the
vertical axis shows the corresponding value of $\mbox{Pr}[n' \!\geq\! n]$.  From
Fig.~\ref{fig:eval_rule_tripadvisor} we could observe that the distribution curve
corresponding to {\em majority rule} lies under the curve corresponding to
{\em average score rule}.  In other words, {\em majority rule} require less ratings
than {\em average score rule}.  This also holds for Fig.~\ref{fig:eval_rule_amazon}
and \ref{fig:eval_rule_ebay}.  These results show that {\em majority rule} is more
robust than {\em average score rule}.

\noindent {\bf Case studies.} We examine the following question:
{\em among TripAdivsor, Amazon and Ebay, whose ratings are more reliable ?}
We answer this question by comparing the distribution of minimum number of ratings
across items for them.  Here we fix the rating aggregation rule to
be {\em majority rule}, since it is more robust than {\em average score rule}.  Again,
we use notation $\mbox{Pr}[n' \!\geq\! n]$ to denote the fraction of items with
the minimum number of ratings larger or equal to $n$.  The numerical results of
$\mbox{Pr}[n' \!\geq\! n]$ are shown in Fig. \ref{fig:reliability_websites},
where the horizontal axis shows the number of ratings $n$, and the vertical axis
shows the corresponding value of $\mbox{Pr}[n' \!\geq\! n]$.  From
Fig. \ref{fig:reliability_websites} one can observe that, as the web sites varies
from Amazon, Ebay, and TripAdvisor, the corresponding distribution curve move up.
In other words, the minimum number of ratings corresponding to Amazon is the
least and that corresponding to TripAdvisor is the highest.  This shows that
Amazon requires less ratings than TripAdvisor and Ebay,
and TripAdvisor requires more ratings than Amazon and Ebay.
Thus, the ratings on Amazon is more reliable than TripAdvisor and Ebay.

Here, we study  the fraction of items (products, hotels, sellers)
which have sufficient number of ratings.
Assume our data sets stated in Table~\ref{data:data_stat} are representative samples
from TripAdvisor, Amazon ane Ebay.  Based on this assumption we  explore,
the fraction of items which have sufficient number of ratings.  We set the
desired minimum number of ratings a product/hotel/seller needs as the median of
the value of minimum number of ratings obtained in the beginning of this section.
Then we test the fraction of products in the data sets stated in
Table~\ref{data:data_stat} satisfy the minimum number of ratings.   Let $f_s$ denote
the fraction of hotels/products/sellers that satisfy the requirement on minimum
number of ratings. The numerical results $f_s$ are shown in
Table \ref{table:fraction_meets_requirement}, where $N_s$ denote the number of
hotels/products/sellers that meet the requirement.  From
Table~\ref{table:fraction_meets_requirement}, we have the following observations:
Only 12.8\% hotels on TripAdvisor, 21.6\% products on Amazon,
74.1\% sellers on Ebay, are with sufficient number of ratings.
Ratings on Ebay is more sufficient than TripAdvisor and Amazon.

\begin{table}
\centering
\begin{tabular}{|c||c|c|c|c|} \hline
&$f_{[0,400]}$ & $f_{(400,800]}$ & $f_{(800,1200]}$ & $f_{>1200}$ \\  \hline \hline
\multicolumn{5}{|c|}{ \bf TripAdvisor }  \\ \hline
{\bf MR}& 39.74\% & 19.05\% & 7.98\% & 33.23\%     \\  \hline
{\bf ASR} &0\% & 0.55\% & 51.9\% & 47.55\%  \\  \hline \hline
\multicolumn{5}{|c|}{ \bf Amazon }  \\ \hline
{\bf MR}& 80.51\% & 7.71\% & 2.75\% & 9.02\%     \\  \hline
{\bf ASR} &0\% & 0.92\% & 37.52\% & 61.56\%  \\  \hline \hline
\multicolumn{5}{|c|}{ \bf Ebay }  \\ \hline
{\bf MR}& 58.06\% & 24.74\% & 5.59\% & 11.61\%     \\  \hline
{\bf ASR} &0\% & 0.03\% & 1.63\% & 98.34\%  \\  \hline
\end{tabular}
\caption{Statistics of minimum number of ratings across hotels in TripAdvisor.}
\label{table:min_num_rat}
\end{table}

\begin{table}[htb]
\centering
\begin{tabular}{|c||c|c|c|c|} \hline
                 & $n'$& \# of items&  $N_s$ & $f_{s}$  \\  \hline \hline
{\bf TripAdvisor}& 573 & 11540 & 1481  & 12.8\%  \\  \hline
{\bf Amazon}     & 146 & 32888 & 7119 & 21.6\%   \\  \hline
{\bf Ebay}       & 356 & 4586 & 3397 & 74.1\% \\  \hline
\end{tabular}
\caption{ Fraction of hotels satisfy the requirement of
minimum number of ratings.}
\label{table:fraction_meets_requirement}
\end{table}

\noindent {\bf Lessons learned \& tips.} Assume we use the
{\em majority rule} to aggregate ratings, we need at least hundreds of ratings for
a hotel (TripAdvisor) / product (Amazon) / seller (Ebay), so as
to produce a reliable evaluation on the quality of a hotel/product or reputation
of a seller.  If the number of ratings of a hotel/product/seller is lower than 100,
then we may be in danger of produce unreliable evaluations.  If we use the
{\em average scoring rule} to aggregate ratings, the system needs more ratings
(around one thousand of ratings for each hotel/product/seller) than using the
{\em majority rule}.  The  {\em majority rule} is more robust and reliable than
the {\em average scoring rule}.  The rating on Amazon is more reliable than
TripAdvisor and Ebay, since to produce a reliable evaluation, Amazon needs less rating
than TripAdvisor and Ebay.  Ratings on Ebay are more
sufficient than TripAdvisor and Amazon.

\subsection{Inferring minimum number of ratings}
In real world applications, it may be expensive or sometimes even impossible to
obtain the exact desired minimum number of ratings for a product.  Here we tackle
this challenge by designing an efficient algorithm to infer the desired
minimum number of ratings, stated in Algorithm~\ref{algo:infer_min_num_rating}.
We show the effectiveness of Algorithm~\ref{algo:infer_min_num_rating} by performing
empirical experiments on TripAdvisor, Amazon and Ebay dataset.

\noindent {\bf Inferring minimum number of ratings algorithm design.}
In real world web services, ratings are with time stamps.
For a given time stamp, we infer the desired minimum number of ratings for a product
based on the historical ratings (up to the given time stamp) of that product.
To infer the desired minimum of ratings, we first infer the model parameter
by applying Algorithm~\ref{infer_paprameter} on the the historical ratings. Then
based on the inferred model parameter, we compute the desired minimum number of ratings.
We outline our algorithm for inferring minimum number of ratings in
Algorithm~\ref{algo:infer_min_num_rating}, where
$\tau_i(t) \!=\! \{\tau_{i,1}, \ldots, \tau_{i, t}\}$
denote a set of historical ratings (up to time stamp $t$) of hotel
$P_i$, and $\tau_{i,1}$ is the earliest rating.
We use $\widehat{n}'$ to denote the inferred minimum number of ratings.

\noindent {\bf Empirical evaluations.} We illustrate the effectiveness of
Algorithm~\ref{algo:infer_min_num_rating} by showing that if a product meets the
inferred minimum number of ratings, then the evaluations of its quality is indeed
reliable.  We work on the same dataset and use the same notations, say
with Section~\ref{section:model_vald}.  The datasets are with time stamps.
For each time stamp, we first infer the minimum number of ratings of that
product by performing Algorithm~\ref{algo:infer_min_num_rating} on its historical
ratings (up to that time stamp).  And then we check whether the number of ratings is
larger than the inferred minimum number of ratings.  If yes, then the number of
test time stamps $N_{test}$ will be increased by one, and at the meantime,
we further evaluate the quality of that product based on its ratings up to that
time stamp, if the evaluated quality reflects the true quality
(obtained by step 3 of Algorithm~\ref{algo:model_valid}), then the
number of reliable evaluations $N_{reliable}$ is increased by one.
We consider {\em majority rule} here.
We perform experiments on TripAdvisor, Amazon and Ebay datasets.  The experiment
results of $N_{test}$ and $N_{reliable}$ are shown
Table~\ref{table:method_application},
where $f_{reliable} = N_{reliable}/N_{test}$.  From Table~\ref{table:method_application} we could observe that if a products meets the requirement on the inferred minimum number of ratings,
then the evaluation of its quality is reliable because the value of
$f_{reliable}$ is 97.27\% for TripAdvisor, 98.50\% for Amazon, 99.56\% for Ebay.
This shows the effectiveness of our algorithm for inferring minimum number of ratings.

\indent {\bf Lessons learned.} Our algorithm of inferring the minimum number of ratings
is effective in producing reliable evaluations of product quality.  Our algorithm
can be deployed into real world applications of web services.

\begin{algorithm}
\caption{Infer minimum number of ratings}
\label{algo:infer_min_num_rating}
\begin{algorithmic}[1]
    \REQUIRE A set of historical ratings $\tau_i(t) \!=\! \{\tau_{i,1}, \ldots, \tau_{i, t}\}$, rating aggregation rule $\SA$, $\delta, Er$
    \ENSURE  Inferred minimum of ratings $\widehat{n}'$.
    \STATE Infer model parameter $\widehat{{\bm \alpha}}_i = \{\widehat{\alpha}_{i,1}, \ldots, \widehat{\alpha}_{i,m}\}$ by performing Algorithm~\ref{infer_paprameter} on the historical rating set $\tau_i(t)$
    \IF {$\SA$ is {\em majority rule}}
        \STATE $\widehat{\alpha}_{i,\ell_i} \leftarrow \max\{\widehat{\alpha}_{i,1}, \ldots, \widehat{\alpha}_{i,m}\}$
        \STATE $\widehat{\alpha}'_{i} \leftarrow \max\{\widehat{\alpha}_{i,1}, \ldots, \widehat{\alpha}_{i,m}\} \setminus \widehat{\alpha}_{i,\ell_i}$
        \STATE $\widehat{n}' \leftarrow \frac{12 \widehat{\alpha}_{i,\ell_i}}{(\widehat{\alpha}_{i,\ell_i} - \widehat{\alpha}'_{i})^2}\ln\frac{m}{\delta}$
    \ENDIF
    \IF {$\SA$ is {\em average score rule}}
        \STATE $\widehat{\gamma}_i \leftarrow \sum_k \widehat{\alpha}_{i,k}$, $\epsilon \leftarrow \frac{-\sqrt{m \widehat{\gamma}_i} + \sqrt{m \widehat{\gamma}_i +4mEr}}{2m}$
        \STATE $\widehat{n}' \leftarrow \frac{3}{\epsilon^2} \ln \frac{2m}{\delta}$
    \ENDIF
    \STATE Compute the minimum number of ratings $\widehat{n}'$ based
\end{algorithmic}
\end{algorithm}

\begin{table}[htb]
\centering
\begin{tabular}{|c||c|c|c|} \hline
                    & $N_{test}$ & $N_{reliable}$ & $f_{reliable}$  \\  \hline \hline
{\bf TripAdvisor} & 907810 & 883030 & 97.27\%   \\  \hline
{\bf Amazon} & 2262036 & 2228173 & 98.50\%   \\  \hline
{\bf Ebay} & 16516962 & 16444179 & 99.56\%   \\  \hline
\end{tabular}
\caption{Fraction of reliable evaluations when products meet the
requirement on inferred minimum number of ratings.} \label{table:method_application}
\end{table}

\section{Related Work } \label{section:related_work}

Several works have investigated the problem of rating aggregation in
online  rating systems.  For example, in
\cite{rep_based_weighted_avg_1, rep_based_weighted_avg_2}, authors
proposed reputation based weighted average score rules that compute
the average rating of products weighted on the reputation score of
users.  Other representative works on rating aggregation can be
found in \cite{ref:rating_agg_ors, prob_based_rating_aggregation_1,
prob_based_rating_aggregation_2, ref:factor_based_rating_agg_ors}.
However, in real online rating system based web services, {\em
average scoring rule} and {\em majority rule} are two most widely
deployed rating aggregation rules, because they are simple, easy to
deploy and more importantly, tested to perform well for many web
services.

Several works have investigated fraud detection.  For example,
authors in \cite{review_spam} explored review spam detection
in online reviewing systems.  In \cite{spam_trad_commu_1,
spam_trad_commu_2} authors studied fraud detection in trading
communities.  Also, fraud detection in recommendation systems was
investigated in \cite{spam_rec_1, spam_rec_2}.  In our paper, we
provide a probabilistic model and formal analysis of users
misbehavior in online ratings systems, and show the condition in
which online rating systems may not be able to reflect the true
quality of products.

Online rating systems are widely deployed in recommendation systems
\cite{rec_sys, ref:recsys_survey}.  Recommendation systems were
introduced since the seminal work on collaborative filtering
\cite{ref:colloab_filt_1, grouplens_1994}. In general,
recommendation systems interpret ratings as preferences of users,
and try to make personalized recommendations by taking into
account the preferences of users.  Researchers investigate various
algorithmic and complexity issues in designing recommendation systems
\cite{rec_sys_eval_coll_fil_algo, rec_sys, ref:recsys_survey}.  The
important distinction of our work is that we treat rating as product
quality assessment and we consider the condition where we
can reveal the true quality with high probability.

\section{ Conclusion} \label{section:conclusion}

We present a mathematical framework to analyze online rating systems that are widely deployed in web services such as TripAdvisor, Amazon, eBay, etc.  We first present a novel mathematical model to specify users' rating behavior.  We then formally analyze a general model of online rating systems.  Through this analysis, we derived theoretical bounds on the desired minimum of the ratings we need to produce a reliable evaluation on the quality of products.  We extend our model to accommodate users' misbehavior (i.e., cheating) in product rating, and show that the {\em majority rule} is more robust and insensitive to misbehavior as compared with the {\em average scoring rule}.  Based on our framework, we also propose an practical algorithm to infer the minimum number of ratings, which is effective in produce reliable product quality evaluations in real-world.  We perform experiments on both synthetic data and three {\em large} real-world data sets (from TripAdvisor, Amazon, Ebay).  Number of interesting observations were found, e.g., generally, hundreds of ratings are sufficient to reflect the true quality of a hotel / product on Amazon / Ebay, or the true reputation of a seller on Ebay.  Ratings on Ebay are more sufficient than TripAdvisor and Amazon.  We believe that our models and methodology can be used as important building blocks to refine and improve online rating systems for various web services.

\bibliographystyle{abbrv}
\bibliography{product_recsys}

\begin{thebibliography}{10}

\bibitem{web:trip_advisor}
Easily rigged travel review sites labeled untrustworthy.
  http://www.tecca.com/columns/travel-review-site-scams-travel-tech/, 2012.

\bibitem{ref:recsys_survey}
G.~Adomavicius and A.~Tuzhilin.
\newblock Toward the next generation of recommender systems: A survey of the
  state-of-the-art and possible extensions.
\newblock {\em IEEE TKDE}, 17(6):734--749, 2005.

\bibitem{ref:prml}
C.~Bishop.
\newblock {\em Pattern recognition and machine learning}.
\newblock springer New York, 2006.

\bibitem{rep_based_weighted_avg_2}
M.~Chen and J.~Singh.
\newblock Computing and using reputations for internet ratings.
\newblock In {\em Proc. of ACM EC}, 2001.

\bibitem{spam_trad_commu_1}
C.~Dellarocas.
\newblock Immunizing online reputation reporting systems against unfair ratings
  and discriminatory behavior.
\newblock In {\em Proc. of ACM EC}, 2000.

\bibitem{gen_prob_limit_theo}
W.~Feller.
\newblock Generalization of a probability limit theorem of cramer.
\newblock {\em Trans. Amer. Math. Soc}, 54(3):361--372, 1943.

\bibitem{rec_sys_eval_coll_fil_algo}
J.~Herlocker, J.~Konstan, L.~Terveen, and J.~Riedl.
\newblock Evaluating collaborative filtering recommender systems.
\newblock {\em ACM TOIS}, 22(1):5--53, 2004.

\bibitem{ref:colloab_filt_1}
W.~Hill, L.~Stead, M.~Rosenstein, and G.~Furnas.
\newblock Recommending and evaluating choices in a virtual community of use.
\newblock In {\em Proc. of ACM CHI}, 1995.

\bibitem{prob_based_rating_aggregation_1}
R.~Jin and L.~Si.
\newblock A study of methods for normalizing user ratings in collaborative
  filtering.
\newblock In {\em Proc. of ACM SIGIR}, 2004.

\bibitem{prob_based_rating_aggregation_2}
R.~Jin, L.~Si, C.~Zhai, and J.~Callan.
\newblock Collaborative filtering with decoupled models for preferences and
  ratings.
\newblock In {\em Proc. of ACM CIKM}, 2003.

\bibitem{review_spam}
N.~Jindal and B.~Liu.
\newblock Analyzing and detecting review spam.
\newblock In {\em Proc. of IEEE ICDM}, 2007.

\bibitem{spam_rec_1}
S.~Lam and J.~Riedl.
\newblock Shilling recommender systems for fun and profit.
\newblock In {\em Proc. of WWW}, 2004.

\bibitem{ref:rating_agg_ors}
H.~W. Lauw, E.-P. Lim, and K.~Wang.
\newblock Quality and leniency in online collaborative rating systems.
\newblock {\em ACM Trans. Web}, 6(1):4:1--4:27, 2012.

\bibitem{book:prob_meth}
J.~Matou\u{s}ek and J.~Vondr\'{a}k.
\newblock {\em The probabilistic method}.

\bibitem{book:prob_comput}
M.~Mitzenmacher and E.~Upfal.
\newblock {\em Probability and computing: Randomized algorithms and
  probabilistic analysis}.
\newblock Cambridge Univiversity Press, 2005.

\bibitem{spam_rec_2}
B.~Mobasher, R.~Burke, and J.~Sandvig.
\newblock Model-based collaborative filtering as a defense against profile
  injection attacks.
\newblock In {\em Proceedings of the National Conference on Artificial
  Intelligence}, 2006.

\bibitem{grouplens_1994}
P.~Resnick, N.~Iacovou, M.~Suchak, P.~Bergstrom, and J.~Riedl.
\newblock Grouplens: an open architecture for collaborative filtering of
  netnews.
\newblock In {\em Proceedings of the ACM conference on Computer supported
  cooperative work}, 1994.

\bibitem{rec_sys}
P.~Resnick and H.~Varian.
\newblock Recommender systems.
\newblock {\em Communications of the ACM}, 40(3):56--58, 1997.

\bibitem{rep_based_weighted_avg_1}
T.~Riggs and R.~Wilensky.
\newblock An algorithm for automated rating of reviewers.
\newblock In {\em Proceedings of the 1st ACM/IEEE-CS joint conference on
  Digital libraries}, pages 381--387. ACM, 2001.

\bibitem{ref:factor_based_rating_agg_ors}
J.~Traupman and R.~Wilensky.
\newblock Collaborative quality filtering: Establishing consensus or recovering
  ground truth?
\newblock {\em Advances in Web Mining and Web Usage Analysis}, pages 73--86,
  2006.

\bibitem{spam_trad_commu_2}
J.~Zhang and R.~Cohen.
\newblock Trusting advice from other buyers in e-marketplaces: the problem of
  unfair ratings.
\newblock In {\em Proc. of ACM EC}, 2006.

\end{thebibliography}

{\section*{Appendix}

\noindent {\large {\bf Proof of Lemma \ref{lemma:rating_dis} }}

Let $\bm \rho \!=\! (\rho_{1}, \ldots, \rho_{m})$ denote the rating
distribution of the user who assigns score $r^+_{i,j}$. Recall that the probability distribution of $\bm \rho $ is $Dirichlet(\bm \alpha_i)$, and $\MP [ r^+_{i,j} \!=\! k | \bm \rho ] \!=\! \rho_{k}$, where $k \!\in\! \{1, \ldots, m\}$. Then we have:
\begin{align*}
&\MP \left[ r^+_{i,j} = k  \right]
  = \int p( \bm \rho ) \MP \left[ r^+_{i,j} = k | {\bm \rho} \right]
    d \bm \rho \\
& \hspace{0.1 in}
  = \int \frac{\Gamma (\sum_{\kappa=1}^m \alpha_{i,\kappa})}{\prod_{\kappa=1}^m
    \Gamma (\alpha_{i,\kappa})}
    \prod_{\kappa=1}^m \rho_{\kappa}^{\alpha_{i,\kappa} - 1}
    \rho_{k} d\rho_{j,1}\ldots d\rho_{j,m} \\
& \hspace{0.1 in}
  = \alpha_{i,k} , \hspace{0.1 in} \mbox{for $k = 1, \ldots, m$},
\end{align*}
so this proof is completed. \done

\noindent {\large {\bf Proof of Theorem \ref{theorem:mr_sincere} }}

Let us state a theorem on bounding the tail probability, that will be used in later derivation.
\begin{theorem}
[Chernoff Bound \cite{book:prob_comput}] Let $X_1, \ldots, X_n$ be
$n$ independent random variables, with $X_i \!=\! 1$ with
probability $p_i$ and 0 otherwise.  Let $X \!=\! \sum_{i=1}^n X_i$
and let $\mu \!=\! E[X]$. Then for each $\epsilon \geq 0$, we have
\begin{align*}
    & \MP[ X \geq (1+\epsilon) \mu ] \leq
      \exp\left(- \min(\epsilon, \epsilon^2) \mu / 3\right), \\
    & \MP[ X \leq ( 1 - \epsilon ) \mu ] \leq
      \exp\left(-\epsilon^2 \mu/ 3  \right).
\end{align*}
\label{theorem:chernoff_bound}
\end{theorem}

%
\vspace{-0.1 in}
Now we can prove Theorem \ref{theorem:mr_sincere} by applying Theorem \ref{theorem:chernoff_bound} as follows.  Let $n_{i,k} \!=\!|\{r_{i,j}| r_{i,j} \in \Sr_i, r_{i,j}=k\}|$ denote the number of ratings that equal to $k$.  By basic probability arguments, we can derive the probability of failing to extract the
true label as follows:
\begin{align}
& \MP \left[ \widehat{\ell}_i \neq \ell_i \right]
  = \MP \left[ \widehat{\ell}_i \neq \ell_i , \:\:
    n_{i, \ell_i} < \frac{ \alpha_{i,\ell_i} + \Sa_i }{2} n_i \right] +  \nonumber \\
&\hspace{0.23 in}
  \MP \left[\widehat{\ell}_i \neq \ell_i, \:\:
      n_{i, \ell_i} \geq \frac{ \alpha_{i,\ell_i} + \Sa_i }{2}n_i \right]  \nonumber \\
&\hspace{0.05 in}
  \leq \MP \left[ n_{i, \ell_i} < \frac{ \alpha_{i,\ell_i} + \Sa_i }{2} n_i \right] +  \nonumber \\
&\hspace{0.23 in}
  \MP \left[ \exists k \neq \ell_i \mbox{ that } n_{i,k} > n_{i,\ell_i}, \:\:
      n_{i, \ell_i} \geq \frac{ \alpha_{i,\ell_i} + \Sa_i }{2}n_i \right]   \nonumber\\
& \hspace{0.05 in}
  \leq \! \MP \!\left[ n_{i, \ell_i} \!<\! \frac{ \alpha_{i,\ell_i} \!\!+\! \Sa_i }{2} n_i \right] \!+\!
    \!\!\sum_{ \begin{subarray}{c}
             k = 1  \\
             k \neq \ell_i
           \end{subarray}
          }
         ^m
         \!\!\MP \!\left[ n_{i,k} \!>\! \frac{ \alpha_{i,\ell_i} \!\!+\! \Sa_i }{2}n_i \right].
\label{ineq:mr_sincere:div_prob}
\end{align}
Let us now proceed to individually derive these two terms of the
above inequality.

One can apply the Chernoff Bound to bound the first term of
Inequality (\ref{ineq:mr_sincere:div_prob}). Observe that $E[
n_{i,k} ] = n_i \alpha_{i,k}$.  Based on this observation, we can
write down the first term of Inequality
(\ref{ineq:mr_sincere:div_prob}) as follows:
\[
\MP\! \left[ n_{i, \ell_i} \!\!<\!\! \frac{ \alpha_{i,\ell_i} \!+\! \Sa_i }{2} n_i \right]
  \!\!=\! \MP \!\left[ n_{i, \ell_i} \! \! <\!\!   E[n_{i, \ell_i}]
    \!\left( 1\!  -\!  \frac{ \alpha_{i,\ell_i}\! -\! \Sa_i }{2\alpha_{i,\ell_i}} \right)  \right],
\]
by applying Chernoff Bound we have
\begin{align*}
 \MP \left[ n_{i, \ell_i} \!\leq\! \frac{ \alpha_{i,\ell_i} \!+\! \Sa_i }{2} n_i \right]
  \! & \leq\! \exp \left(\! -
     \frac{(\alpha_{i,\ell_i} \!-\! \Sa_i)^2}{3(2\alpha_{i,\ell_i})^2}
    E[ n_{i, \ell_i } ] \!\right) \\
  & = \exp \left(\! -
     \frac{(\alpha_{i,\ell_i} \!-\! \Sa_i)^2}{12\alpha_{i,\ell_i}}
     n_{i } \!\right),
\end{align*}
by substituting $n_i$ with Inequality
(\ref{ineq:mr_sincere:rating_bound}), we have
\begin{equation}
    \MP \left[ n_{i, \ell_i} \!\leq\! \frac{ \alpha_{i,\ell_i} \!+\! \Sa_i }{2} n_i \right]
    \leq \frac{\delta}{m}.
\label{ineq:mr_sincere:first_term}
\end{equation}
The remaining issue is to derive the last term of Inequality
(\ref{ineq:mr_sincere:div_prob}).  Since $E[ n_{i,k} ] = n_i
\alpha_{i,k}$. Based on this observation, we can write down the last
term of Inequality (\ref{ineq:mr_sincere:div_prob}) as follows:
\begin{align}
& \sum_{ \begin{subarray}{c}
             k = 1  \\
             k \neq \ell_i
           \end{subarray}
          }
         ^m
  \MP \left[ n_{i,k} > \frac{ \alpha_{i,\ell_i} + \Sa_i }{2}n_i \right]  \nonumber \\
& \hspace{0.1 in}
  = \sum_{ \begin{subarray}{c}
             k = 1  \\
             k \neq \ell_i
           \end{subarray}
          }
         ^m
  \MP \left[ n_{i,k} > E[ n_{i,k} ] \left( 1 + \frac{ \alpha_{i,\ell_i} +
  \Sa_i - 2\alpha_{i,k} }{2\alpha_{i,k}} \right) \right]  \nonumber \\
& \hspace{0.1 in}
  \leq \sum_{ \begin{subarray}{c}
             k = 1  \\
             k \neq \ell_i
           \end{subarray}
          }
         ^m
  \exp \left( - \frac{ E[ n_{i,k} ] }{3}  \times \right.  \nonumber \\
& \hspace{0.18 in}
\left. \mbox{min} \left\{
  \frac{ \alpha_{i,\ell_i} + \Sa_i - 2\alpha_{i,k} } {2\alpha_{i,k}},
  \frac{(\alpha_{i,\ell_i} + \Sa_i - 2\alpha_{i,k} )^2 }{ ( 2\alpha_{i,k} )^2}
  \right\}   \right),
\label{ineq:mr_sincere:last_term}
\end{align}
where the last step is obtained by applying Chernoff Bound. Since
$\alpha_{i,\ell_i} + \Sa_i - 2\alpha_{i,k} \geq \alpha_{i,\ell_i} -
\Sa_i \geq 0$, thus we have
\begin{align*}
& \mbox{min} \left\{
  \frac{ \alpha_{i,\ell_i} + \Sa_i - 2\alpha_{i,k} } {2\alpha_{i,k}},
  \frac{(\alpha_{i,\ell_i} + \Sa_i - 2\alpha_{i,k} )^2 }{ ( 2\alpha_{i,k} )^2}
  \right\}   \\
& \hspace{0.1 in}
  \geq \mbox{min} \left\{
  \frac{ \alpha_{i,\ell_i} - \Sa_i } {2\alpha_{i,k}},
  \frac{(\alpha_{i,\ell_i} - \Sa_i )^2 }{ ( 2\alpha_{i,k} )^2}
  \right\}  \\
& \hspace{0.1 in}
  \geq \frac{ \alpha_{i,\ell_i} - \Sa_i } {2\alpha_{i,k}}
  \mbox{min} \left\{ 1, \frac{ \alpha_{i,\ell_i} - \Sa_i } {2\Sa_i}
  \right\}, \hspace{0.1 in} \forall k \neq \ell_i,
\end{align*}
applying this inequality to Inequality
(\ref{ineq:mr_sincere:last_term}), we obtain
\begin{align}
& \sum_{ \begin{subarray}{c}
             k = 1  \\
             k \neq \ell_i
           \end{subarray}
          }
         ^m
  \MP \left[ n_{i,k} > \frac{ \alpha_{i,\ell_i} + \Sa_i }{2}n_i \right]  \nonumber \\
& \hspace{0.1 in}
  \leq \sum_{ \begin{subarray}{c}
             k = 1  \\
             k \neq \ell_i
           \end{subarray}
          }
         ^m
  \exp \left( - \frac{ ( \alpha_{i,\ell_i} - \Sa_i ) n_i }{6}
  \mbox{min} \left\{ 1, \frac{ \alpha_{i,\ell_i} - \Sa_i } {2\Sa_i}
  \right\} \right).
\label{ineq:mr_sincere:last_term_final}
\end{align}
We can further simplify this inequality under the following two cases. \\
{\bf Case 1:} $\Sa_i \!< \!\alpha_{i, \ell_i} / 3$. In this case,
$(\alpha_{i,\ell_i}\! -\! \Sa_i ) /(2\Sa_i) \!>\! 1$.  Therefore
$\mbox{min} \left\{ 1, ( \alpha_{i,\ell_i} - \Sa_i ) / (2\Sa_i)
\right\}\! =\! 1$. Applying this to Inequality
(\ref{ineq:mr_sincere:last_term_final}) and substituting $n_i$ with
Inequality (\ref{ineq:mr_sincere:rating_bound}), we have
\begin{align}
& \sum_{ \begin{subarray}{c}
             k = 1  \\
             k \neq \ell_i
           \end{subarray}
          }
         ^m
\MP \! \left[ n_{i,k}\!  >\!  \frac{ \alpha_{i,\ell_i} + \Sa_i }{2}n_i \right]
\! \leq\!
  \sum_{ \begin{subarray}{c}
             k = 1  \\
             k \neq \ell_i
           \end{subarray}
          }
         ^m
  \exp \left( - \frac{2 \alpha_{i,\ell_i}}{ \alpha_{i,\ell_i} - \Sa_i }
   \ln \frac{m}{\delta} \right)  \nonumber \\
&\hspace{0.5 in} \leq
  \sum_{ \begin{subarray}{c}
             k = 1  \\
             k \neq \ell_i
           \end{subarray}
          }
         ^m
   \exp \left( - 2 \ln \frac{m}{\delta} \right)  = \frac{m-1}{m^2} \delta^2.
\label{ineq:mr_sincere:case1}
\end{align}
{\bf Case 2:} $\Sa_i \geq \alpha_{i, \ell_i} / 3$.  In this case,
$(\alpha_{i,\ell_i} - \Sa_i ) /(2\Sa_i) \leq 1$.  We have
$    \mbox{min} \left\{ 1, ( \alpha_{i,\ell_i} - \Sa_i ) / (2\Sa_i)
\right\} =
    (\alpha_{i,\ell_i} - \Sa_i ) /(2\Sa_i)$.
Applying this equation to Inequality
(\ref{ineq:mr_sincere:last_term_final}) and substituting $n_i$ with
Inequality (\ref{ineq:mr_sincere:rating_bound}), we have
\begin{align}
& \sum_{ \begin{subarray}{c}
             k = 1  \\
             k \neq \ell_i
           \end{subarray}
          }
         ^m
  \MP \left[ n_{i,k} > \frac{ \alpha_{i,\ell_i} + \Sa_i }{2}n_i \right]
\leq
  \sum_{ \begin{subarray}{c}
             k = 1  \\
             k \neq \ell_i
           \end{subarray}
          }
         ^m
  \exp \left( - \frac{\alpha_{i,\ell_i}}{ \Sa_i }
   \ln \frac{m}{\delta} \right)  \nonumber \\
&\hspace{0.5 in} \leq
  \sum_{ \begin{subarray}{c}
             k = 1  \\
             k \neq \ell_i
           \end{subarray}
          }
         ^m
   \exp \left( - \ln \frac{m}{\delta} \right) = \frac{m-1}{m} \delta.
\label{ineq:mr_sincere:case2}
\end{align}
Combining Inequality (\ref{ineq:mr_sincere:case1}) and Inequality
(\ref{ineq:mr_sincere:case2}), we have
\begin{equation}
   \sum_{ \begin{subarray}{c}
             k = 1  \\
             k \neq \ell_i
           \end{subarray}
          }
         ^m
  \MP \left[ n_{i,k} > \frac{ \alpha_{i,\ell_i} + \Sa_i }{2}n_i \right]
  \leq \frac{m - 1}{m} \delta.
\label{ineq:mr_sincere:last_term_bound}
\end{equation}
Applying Inequality (\ref{ineq:mr_sincere:first_term}) and
Inequality (\ref{ineq:mr_sincere:last_term_bound}) to Inequality
(\ref{ineq:mr_sincere:div_prob}), we complete the proof. \done

\noindent {\large {\bf Proof of Theorem
\ref{theorem:mr_sincere:tightness}}}

Let us state a theorem on bounding the tail probability, that will be used in later derivation.
\begin{theorem}
[\cite{book:prob_meth, gen_prob_limit_theo}] Let $X$ be a sum
of independent random variables, each attaining values in $[0, 1]$,
and let $\sigma = \sqrt{ \MV [ X ] } \geq 200$. Then for all $t \in
[0, \frac{\sigma^2}{100}]$, we have
\[
\MP [ X \geq E[X] + t] \geq c \exp \left( - \frac{ t^2 }{3 \sigma^2} \right),
\]
for a suitable constant $c > 0$.
\label{theorem:mr_sincere:low_bound_binomial}
\end{theorem}

Now we can prove Theorem \ref{theorem:mr_sincere:tightness} by applying Theorem \ref{theorem:mr_sincere:low_bound_binomial} as
follows.  Recall that $n_{i,k}$ is the number of ratings that equal
$k$. Let $\widetilde{\ell}_i = \arg\min_{k }\{ |\alpha_{i,k} -
\widetilde{\alpha}_i|\}$.  We can derive a general lower bound of
the probability that we fail to extract the true label as follows:
\[
    \MP \left[ \widehat{\ell}_i \neq \ell_i \right] \geq
    \MP \left[ n_{i, \widetilde{\ell}_i} > n_{i, \ell_i} \right].
\]
Let $R_{i,j}$, $j\! \in\! \{1,\ldots,n_i \}$ denote a set of random
variables with
\[
R_{i,j} = \left\{
\begin{aligned}
   & 1, \hspace{0.1 in} && \mbox{with probability } \MP [ r^+_{i,j} = \widetilde{\ell}_i ] \\
   & 0, \hspace{0.1 in} &&\mbox{with probability } \MP [ r^+_{i,j} = \ell_i ] \\
   & 1/2, \hspace{0.1 in} && \mbox{otherwise}\\
\end{aligned}
\right. ,
\]
where the probability mass function of $r^+_{i,j}$ is derived in
Lemma \ref{lemma:rating_dis}.  Observe that $R_i \!>\! n_i / 2$ if
and only if the number of ratings equal $\ell_i$ is smaller than the
number of ratings equal $\widetilde{\ell}_i$, or $n_{i,
\widetilde{\ell}_i} \!>\! n_{i, \ell_i}$. Thus we have $ \MP \left[
\widehat{\ell}_i \!\neq\! \ell_i \right] \!\geq\! \MP \left[ R_i
\!>\! \frac{n_i}{2} \right]$. In the following, we seek to complete
the proof by showing $\MP [ R_i \!>\! n_i / 2 ] \geq
\Omega(\delta)$.

Here we apply Theorem \ref{theorem:mr_sincere:low_bound_binomial} to
give a lower bound of $\MP [ R_i > n_i / 2 ]$.  First we can express
the expectation and variance of $R_i$ as:
\begin{align*}
  & E[ R_i ] = n_i / 2 -
     ( \alpha_{i, \ell_i} - \widetilde{\alpha}_i ) n_i / 2 ,   \\
  & \MV [ R_i ] =  \left( \alpha_{i, \ell_i} + \widetilde{\alpha}_i -
    ( \alpha_{i, \ell_i} - \widetilde{\alpha}_i )^2 \right)n_i / 4.
\end{align*}
Let $t = n_i ( \alpha_{i, \ell_i} - \widetilde{\alpha}_i ) / 2 +
\epsilon$, where $0< \epsilon<10^{-10}$.  Observe that $E[R_i] +t =
n_i / 2 +\epsilon > n_i / 2$, thus $\MP \left[ R_i > \frac{n_i}{2}
\right] \geq \MP \left[ R_i \geq E[R_i] +t \right]$.  Before
applying Theorem \ref{theorem:mr_sincere:low_bound_binomial} to
derive a lower bound of $\MP \left[ R_i \!\geq\! E[R_i] +t \right]$,
we need to check whether some conditions specified in Theorem
\ref{theorem:mr_sincere:low_bound_binomial} are satisfied.

Let us first check whether $\sqrt{\MV [ R_i ]} \geq 200$ holds.
Since
\begin{align}
& \frac{100}{101}\alpha_{i, \ell_i} \! \leq\!  \widetilde{\alpha}_i
 \Leftrightarrow
  \alpha_{i, \ell_i} - \widetilde{\alpha}_i \leq
  \frac{1}{201} ( \alpha_{i, \ell_i} + \widetilde{\alpha}_i ).
\label{ineq:mr_sincere:tightness_1}
\end{align}
Based this observation, we can derive a lower bound of $\alpha_{i,
\ell_i} + \widetilde{\alpha}_i - ( \alpha_{i, \ell_i} -
\widetilde{\alpha}_i )^2$ as follow:
\begin{align}
& \alpha_{i, \ell_i} + \widetilde{\alpha}_i -
  ( \alpha_{i, \ell_i} - \widetilde{\alpha}_i )^2
  \geq
  \alpha_{i, \ell_i} + \widetilde{\alpha}_i -
  \left( \frac{\alpha_{i, \ell_i} + \widetilde{\alpha}_i}{201} \right)^2
  \nonumber \\
& \hspace{0.2 in}
  \geq \frac{200 \times 202}{201^2} \left(\alpha_{i, \ell_i} +
  \frac{100}{101} \alpha_{i, \ell_i} \right)  \geq 1.99 \alpha_{i, \ell_i}.
\label{ineq:mr_sincere:tightness_2}
\end{align}
By applying Inequality (\ref{ineq:mr_sincere:tightness_2}), the
lower bound of $\MV [ R_i ]$ is
\begin{align*}
\MV [ R_i ]
& \geq \frac{1.99}{4} n_i \alpha_{i, \ell_i}
= O \left( \frac{1.99}{4} \frac{\alpha^2_{i, \ell_i}}{( \alpha_{i, \ell_i} -
   \widetilde{\alpha}_i )^2} \ln \frac{1}{\delta} \right) \\
& \geq O \left( \frac{1.99\times 101^2}{4} \ln \frac{1}{\delta}   \right).
\end{align*}
By making the left side of this inequality larger or equal to 40000,
we guarantee that $\sqrt{ \MV [ R_i ]} \!\geq\!  200 $, or $\delta
\!\leq\! O ( e^{-7.9})$.  Hence, by letting $\eta \!=\! O (
e^{-7.9})$, then for any $\delta \!\leq\! \eta$, we have $\sqrt{ \MV
[ R_i ]} \!\geq\! 200 $.

Let us continue to check whether $t \in [ 0, \MV [ R_i ] / 100 ]$.
By apply Inequality (\ref{ineq:mr_sincere:tightness_1}), we could
derive an upper bound of $ t /\MV [ R_i ] $ as follows:
\begin{align*}
\frac{ t }{\MV [ R_i ]}
& = \frac{2 (\alpha_{i, \ell_i} - \widetilde{\alpha}_i) }
         { \alpha_{i, \ell_i} + \widetilde{\alpha}_i -
           ( \alpha_{i, \ell_i} - \widetilde{\alpha}_i )^2
         } + \frac{ \epsilon }{\MV [ R_i ]}  \\
& \leq 2 \left( \frac{ \alpha_{i, \ell_i} + \widetilde{\alpha}_i }
                     { \alpha_{i, \ell_i} - \widetilde{\alpha}_i } -
                (\alpha_{i, \ell_i} - \widetilde{\alpha}_i )
         \right)^{-1} + 10^{-14}\\
& \leq 2 \left( 201 - \frac{1}{201}( \alpha_{i, \ell_i} + \widetilde{\alpha}_i )
         \right)^{-1} + 10^{-14} \leq \frac{1}{100}.
\end{align*}
Thus we have that $t \in [ 0, \MV [ R_i ] / 100 ]$.

Finally, we can apply Theorem
\ref{theorem:mr_sincere:low_bound_binomial} to derive a lower bound
of $\MP \left[ R_i \!>\! n_i / 2 \right]$ as follows:
\begin{align*}
& \MP \left[ R_i \!>\! \frac{n_i}{2} \right]
  \geq \MP [ R_i \geq E [ R_i ] + t ]
  \geq c \exp \left( - \frac{ t^2 }{3 \MV [ R_i ]} \right) \\
& \hspace{0.1 in}
  = c \exp \left( - \frac{1}{3} \frac{ n_i ( \alpha_{i, \ell_i} -
   \widetilde{\alpha}_i )^2 }{ \alpha_{i, \ell_i} + \widetilde{\alpha}_i -
   ( \alpha_{i, \ell_i} - \widetilde{\alpha}_i )^2 } \right)\\
& \hspace{0.1 in}
  \geq c \exp \left( - \frac{1}{3} \frac{ n_i ( \alpha_{i, \ell_i} -
   \widetilde{\alpha}_i )^2 }{1.99 \alpha_{i, \ell_i} } \right)
  \geq \Omega ( \delta )
\end{align*}
Thus, $ \MP \left[ \widehat{\ell}_i \neq \ell_i \right] \geq \MP
\left[ R_i > \frac{n_i}{2} \right] \geq \Omega ( \delta ) $. \done

\noindent {\large {\bf Proof of Theorem \ref{theorem:mr_random_misb}
} }

We prove this theorem by extending Theorem \ref{theorem:mr_sincere}.
First let us present a probabilistic interpretation of the random
misbehaving users as follows: with probability $f_i$ we generate a
misbehaving user to rate product $P_i$, and with probability $1 -
f_i$ we generate an honest user to rate product $P_i$.  This
interpretation describes a new generative process for ratings with
random misbehaving users.  Let $r_i$ be a rating generated by that
new generative process.  Observe that $\MP [ r_i = k | \mbox{random
misbehavior} ] = \frac{1}{m}$ and $\MP [ r_i = k | \mbox{sincere} ]
= \alpha_{i,k}$, where $k = 1, \ldots, m$.  Then we can compute the
probability mass function of $r_i$ as follows:
\[
    \MP [ r_i = k ] = \frac{f_i}{m} + (1-f_i)\alpha_{i,k},
    \hspace{0.1 in}
    \mbox{ for $k = 1, \ldots, m$}.
\]
It follows that $\MP [ r_i \!=\! \ell_i ] \!=\! \arg\max\{\MP [ r_i
\!=\! k ] \}$ and $f_i / m +(1-f_i)\widetilde{\alpha}_i$ are the
largest and second largest among $\MP [ r_i \!=\! 1 ], \ldots, \MP [
r_i \!=\! m ]$ respectively. Then, by a similar derivation with
Theorem \ref{theorem:mr_sincere}, we finish this proof. \done

%
%
\noindent {\large {\bf Proof of Theorem
\ref{theorem:mr_bias:bias_win}} }

We prove this theorem by extending Theorem \ref{theorem:mr_sincere}.
First let us present a probabilistic interpretation of the bias
misbehaving users as follows: with probability $f'_i$ we generate a
biased misbehaving user to rate product $P_i$, and with probability
$1 - f'_i$ we generate a honest user to rate product $P_i$.  This
interpretation describes a new generative process for ratings with
biased misbehaving users. Let $r'_i$ be a rating generated by that
new generative process. Observe that $\MP [ r'_i \!=\! k |
\mbox{biased misbehavior} ] \!=\! 1$ with $k \!=\! \ell'_i$ and 0
otherwise, and $\MP [ r'_i \!=\! k | \mbox{sincere} ] \!=\!
\alpha_{i,k}, \forall k$. Then we have:
\[
 \MP \left[ r'_i = k \right] = \left\{
 \begin{aligned}
    & f'_i + (1-f'_i)\alpha_{i, \ell'_i},
      && \hspace{0.1 in } \mbox{for $k = \ell'_i$}  \\
    & (1-f'_i)\alpha_{i,k}, && \hspace{0.1 in } \mbox{otherwise}
 \end{aligned}
\right. .
\]
We have two cases to consider:\\
{\bf Case 1:} $\ell'_i \neq \ell_i$. Observe that
\[
    f'_i > \frac{ \alpha_{i,\ell_i} - \alpha_{i, \ell'_i} }
           { 1 + \alpha_{i,\ell_i} - \alpha_{i, \ell'_i} }
    \Leftrightarrow
    f'_i + (1-f'_i)\alpha_{i,\ell'_i} > (1-f'_i)\alpha_{i,\ell_i}.
\]
Thus $\MP [ r'_i \!=\! \ell'_i ] \!=\! f'_i + (1-f'_i)\alpha_{i,
\ell'_i}$ and $\MP [ r'_i \!=\! \ell_i ] \!=\!
(1-f'_i)\alpha_{i,\ell_i}$ are the largest and second largest among
$\MP [ r'_i \!=\! 1 ], \ldots, \MP [ r'_i \!=\! m ]$ respectively.
Therefore, with a large enough value of $n_i$, the number of ratings
that equal to $\ell'_i$ will be the largest with high probability.
Then by a similar derivation with Theorem \ref{theorem:mr_sincere},
we could finish the proof for this case.

\noindent {\bf Case 2:} $\ell'_i \!=\! \ell_i$.  Observe that $\MP [
r'_i \!=\! \ell_i ] \!=\! f'_i + (1 - f'_i)\alpha_{i, \ell_i}$ and
$(1-f'_i)\widetilde{\alpha}_i$ are the largest and second largest
among $\MP [ r' \!=\! 1 ], \ldots, \MP [ r' \!=\! m ]$ respectively.
Thus we extract the label $\ell_i$ with a high probability with
large enough value of $n_i$. Finally, by a similar derivation with
Theorem \ref{theorem:mr_sincere}, one completes the proof.  \done

\subsection{ {\bf Proof of Theorem \ref{theorem:as_sincere} } }

Let $\beta_{i,k} \!=\! \sum_{j=k}^{m} \alpha_{i,j}$. Observe that $\gamma_i
\!=\! \sum_{k=1}^m k \alpha_{i,k} \!=\! \sum_{k=1}^m \beta_{i,k}$. Recall that
we use $n_{i,k}$, $k = 1, \ldots, m$, to denote the number of ratings that
equal to $k$.  Let $n'_{i,k} = \sum_{j = k}^m n_{i,j}$, $k = 1, \dots, m$,
denote the number of ratings which are no less than $k$. Then we have
\[
    \sum_{j=1}^{M} r_{i,j} = \sum_{k=1}^m k n_{i,k} = \sum_{k=1}^m n'_{i,k}.
\]
Then it follows that our problem can be formulated to bound
\[
    \left| \widehat{r}_i - \gamma_i \right| =
    \left| \sum_{k=1}^m \frac{ n'_{i,k} }{n_i} - \sum_{k=1}^m \beta_{i,k} \right|.
\]
In the following, we will first seek to bound $| n'_{i,k} / n_i - \beta_{i,k}
|$, and then we apply this bound to complete our proof.

We bound $| n'_{i,k} / n_i - \beta_{i,k} |$ by the following claim:
\begin{itemize}
   \item {\bf Claim 1:} For each $k \in \{1, \ldots, m\}$, if Inequality
       (\ref{ineq:asr_sincere:ni}) holds, then
    \begin{equation}
    \left| \frac{n'_{i,k}}{n_i} - \beta_{i,k} \right| \!\leq\! \left\{
      \begin{aligned}
         & \epsilon^2, \hspace{0.1 in} \beta_{i,k} \leq \epsilon^2 \\
         & \epsilon \sqrt{\beta_{i,k}}, \hspace{0.1 in} \mbox{otherwise}
      \end{aligned}
        \right.
    \label{ineq:as:bound_prob}
    \end{equation}
    holds with probability at least $1 - \delta/m$.
\end{itemize}
Let us prove this claim here.  Let $\mathbf{I}_{j,k}$, $j=1, \ldots, n_i$, $k =
1, \ldots, m$ be a set of indicator random variables such that
\[
    \mathbf{I}_{j,k} \!=\! \left\{
    \begin{aligned}
       & 1 , \hspace{0.1 in} \mbox{ if $ r^+_{i,j} \geq k$ } \\
       & 0 , \hspace{0.1 in} \mbox{otherwise}
    \end{aligned}
    \right. .
\]
From Lemma \ref{lemma:rating_dis} we have that $\MP [ r^+_{i,j} \!=\! k ] \!=\!
\alpha_{i,k}$, where $k \!=\! 1, \ldots, m$ and $j \!=\! 1, \ldots, n_i$. Thus
$\MP [ r^+_{i,j} \geq k ] \!=\! \beta_{i,k}$.  Therefore, we could have $\MP [
\mathbf{I}_{j,k} = 1 ] \!=\! \beta_{i,k} $.  It follows that  $E \left[
\sum_{j=1}^{n_i} \mathbf{I}_{j,k} \right] \!=\! n_i \beta_{i,k} $.  Observe
that $n'_{i,k} = \sum_{j=1}^{n_i} \mathbf{I}_{j,k}$. Let us consider a special
case, $\beta_{i,k} \!=\! 0$ first.  For this special case, by a simple check,
we could see that claim 1 holds. Then we consider the case, $\beta_{i,k} \geq
\epsilon^2$.  Observe that for this case the inequality $0 < \epsilon /
\sqrt{\beta_{i,k}} < 1$ holds. Based on this observation, we can further show:
\begin{align*}
& \MP \left[ \left| \frac{n'_{i,k}}{n_i} - \beta_{i,k} \right| \leq \epsilon \sqrt{\beta_{i,k}} \right] \\
& \hspace{0.1 in}
  \geq 1 - \MP \left[ \left| \frac{n'_{i,k}}{n_i} - \beta_{i,k} \right|
       \geq \epsilon \sqrt{\beta_{i,k}} \right] \\
& \hspace{0.1 in}
  = 1 - \MP \left[ \left| \sum_{j=1}^{n_i} \mathbf{I}_{j,k} - n_i \beta_{i,k} \right|
   \geq \frac{\epsilon }{\sqrt{\beta_{i,k}}} n_i \beta_{i,k} \right] \\
& \hspace{0.1 in}
  \geq 1 - 2 \exp \left( - n_i \frac{\epsilon^2}{3} \right)
  \geq 1 - \frac{\delta}{m},
\end{align*}
where the last two steps are obtained by applying Chernoff Bound and by
substituting $n_i$ with Inequality (\ref{ineq:asr_sincere:ni}) respectively.
Finally, we consider the case, $0 < \beta_{i,k} < \epsilon^2 $.  Observe that
for this case the inequality $\epsilon / \sqrt{\beta_{i,k}} > 1$ holds. Then by
a derivation as the previous case, we obtain
\[
    \MP \left[ \left| \frac{n'_{i,k}}{n_i} - \beta_{i,k} \right| \leq \epsilon^2 \right] \geq 1 - \frac{\delta}{m}
\]
holds.  Hence we proved Claim 1.

Let $E_k$ denote the event that Inequality (\ref{ineq:as:bound_prob}) holds for
a specific value $k$.  Then based on claim 1, we have $\MP [ E_k ] \geq \delta
/ m$.  Then we can derive the probability that events $E_1, \ldots, E_m$ all
holds as follows:
\begin{align*}
\MP \left[ \bigcap_{k=1}^m E_k \right]
& = 1 - \MP \left[ \bigcup_{k=1}^m \overline{E}_k  \right]
 \geq 1 - \sum_{k=1}^m \MP \left[\: \overline{E}_k \right] \\
& \geq 1 - \sum_{k=1}^m \frac{\delta}{m}
 \geq 1 - \delta.
\end{align*}
Suppose that events $E_1, \ldots, E_m$ all holds then,
\begin{align*}
\left| \widehat{r}_i - \gamma_i \right|
& =
  \left| \sum_{k=1}^m \frac{n'_{i,k}}{n_i} - \sum_{k=1}^m \beta_{i,k} \right| \\
& \leq \sum_{k=1}^m \left| \frac{n'_{i,k}}{n_i} - \beta_{i,k} \right| \\
& \leq \epsilon \sum_{k=1}^m \mbox{max}\left\{ \sqrt{\beta_{i,k}},
  \epsilon \right\}  \\
& \leq \epsilon \sum_{k=1}^m \sqrt{\beta_{i,k}}+ m \epsilon^2
\end{align*}
by applying Cauthy's Inequality we have:
\begin{align*}
\left| \widehat{r}_i - \gamma_i \right|
& \leq  \epsilon \sqrt{ m \sum_{k=1}^m \beta_{i,k} }  + m \epsilon^2 \\
& = \epsilon \sqrt{m \gamma_i} + m \epsilon^2
\end{align*}
The proof can be completed by recalling that the probability that events $E_1,
\ldots, E_m$ all holds is at least $1 - \delta$. \done

\subsection{ {\bf Proof of Theorem \ref{theorem:as_random_misb} }}

When there is $f_i$ fraction of random misbehaving users, the average rating
$\widehat{r}_i$ of product $P_i$ converges to $(1 - f_i)\gamma_i + f_i m/2
\!=\! \gamma_i + ( m/2 - \gamma_i )f_i$. Since $n_i \geq 3 \ln(2m/\delta) /
\epsilon^2$ holds, then with a similar derivation of Theorem
\ref{theorem:as_sincere}, we could obtain that:
\begin{align}
& \left| \sum\nolimits_j r_{i,j} / n_i - \left(\gamma_i + ( m / 2 - \gamma_i ) f_i \right) \right| \nonumber \\
& \hspace{0.1 in}  \leq
    \epsilon \sqrt{ m \left(\gamma_i + ( m / 2 - \gamma_i ) f_i \right) } + m \epsilon^2
\label{ineq:as_random_misb_proof}
\end{align}
holds with probability at least $1 -\delta$. Observe that
\begin{align*}
& \left| \frac{m}{2} \!-\! \gamma_i \right| f_i \!-\! \left| \frac{\sum_j r_{i,j} }{n_i} \!-\!
  \left( \gamma_i \!+\! \frac{m}{2}f_i - \gamma_i f_i \right) \right|
  \!\leq\! \left| \frac{\sum_j r_{i,j} }{n_i} \!-\! \gamma_i \right| \\
&\hspace{0.5 in}
  \leq \left| \frac{m}{2} - \gamma_i \right| f_i +
  \left| \frac{\sum_j r_{i,j} }{n_i} -
  \left( \gamma_i + \frac{m}{2}f_i - \gamma_i f_i \right) \right|,
\end{align*}
by applying Inequality (\ref{ineq:as_random_misb_proof}) to this inequality we
can finish the proof of this theorem.  \done
}

\end{document}